\documentclass[]{Tauola_interface_design}
\usepackage{graphicx}
\usepackage{hyperref}
\usepackage{eurosym}
\usepackage{pifont}
\usepackage{amsmath}
\usepackage{subfigure}

\makeatletter
\renewcommand*\l@subsection{\@dottedtocline{2}{1.5em}{2.0em}}
\renewcommand*\l@subsubsection{\@dottedtocline{3}{3.0em}{3.0em}}
\makeatother

\begin{document}

\maketitle

\tableofcontents\pdfbookmark[0]{Table of Contents}{toc}

\newpage

\section{Introduction}
Before we will present our C++ Interface of TAUOLA,
for LHC era generators, let us first list projects and applications which helped us
to gain experience and to define the main tasks our interface 
for a $\tau$ decay generator has to resolve.

In the present day experiments at High Energy Physics accelerators, 
interpretation of  results becomes increasingly involved. 
Not only the detector 
response is complex, but some theoretical effects need to be 
removed. Otherwise results are difficult to interpret for the non-specialist. 
For that purpose the concept of work with realistic and idealized observables was  introduced as well 
as finally with pseudo-observables which can be easily
understood by theorists, such as $W$, $Z$ masses  or couplings.
Good examples of 
such approaches  were measurements of the two-fermion final states at LEP.  
Because of increasing  precision of the experimental measurements, definitions of 
quantities to be measured, simple at first, later evolved into 
several options \cite{Kobel:2000aw},
each based on the properties of individual detectors and each requiring individual discussion of the systematic 
error.
One could assume, that if all  theoretical effects are embodied into one 
theoretical black-box 
and, experiments while using it tune parameters (representing pseudo-observables) to the data, interpretation of the observed effects
could be separated into theoretical and experimental components.
Unfortunately this strategy is limited, as it leaves little room 
for cases where 
theory and experimental effects are convoluted: size and even nature of the theoretical corrections 
depend on the experimental conditions. Such discussion on observables involving $\tau$ decays can be found in \cite{RichterWas:2004jf}.
Recently,  discussion for the physics of 
$\tau$ production and decay at low energies where similar aspects are addressed,  
was presented in \cite{Actis:2009gg}. 

For LHC experiments $\tau$ 
decays are not of primary interest in themselves, but rather 
will be  used to measure properties of $\tau$ production processes. 
Let us explain this using examples. Physics effects necessary for 
the prediction of hard processes 
at the LHC experiments can be separated into several parts, among them: 
parton showers, 
the underlying event and structure functions, final 
state QED bremsstrahlung, QED bremsstrahlung interference between initial and 
final states and finally the hard process including electroweak corrections. 
Such separation 
is not only for the convenience of organizing theoretical work but 
provides efficient and flexible
   component in the framework for experimental data analysis
 strategy (see eg. \cite{Aad:2009wy}).
Some
of such building blocks are of genuine theoretical interests, some others
are not so much. The
 hard process
usually  depends on the parameters intended for the measurement, eg.  
$W$ or Higgs mass, or new coupling constants. 
Other building blocks may be less interesting, nonetheless they may affect
the results of measurements. This is certainly true in the case of the underlying event
or missing transverse energy or $p_T$ distributions generated from parton 
showers \cite{RichterWas:2004jf}. It may also be the case for QED final
state bremsstrahlung or
initial-final state interference (where potential difficulties may be expected 
\cite{Abdallah:2005wn} and predictions may need to be fixed with the 
help of experimental data).

The black-box approach, where all simulation segments are put together by
theorists, may look advantageous to the experimental user.
However in such a case one has less flexibility to distinguish experimental
effect from the theoretical ones, thus limiting control on the systematic errors. 
The particular problems may be left unnoticed. Typically
 the difficulties will not affect all
observables. Unfortunately, complications tend to show up only when more 
detailed discussion on the systematic errors of  experimental analysis is
performed. 


In the present document we discuss the implementation of $\tau$ decays 
into a simulation chain as a separate module which can be configured by the end 
user. For the purpose of generation of $\tau$ decays themselves, 
the {\tt TAUOLA}  library, as described
in \cite{Jadach:1990mz,Jezabek:1991qp,Jadach:1993hs} 
is used. This part of the code is expected to be
a black-box for the High Energy experimental user.  
At present, from the technical side, the black-box consists of 
 the same {\tt FORTRAN} code  as described 
in \cite{Golonka:2003xt}. We will call it {\tt TAUOLA FORTRAN}. Such organization
 makes it
 easy for low energy phenomenologists or experimentalists to work on this part of the code,
such as the activities described in \cite{Actis:2009gg}, leading to the
new parametrization of hadronic $\tau$ decay currents  becoming available for  
High Energy experimental users in a rather straightforward way.

The role of the interface is to prepare information on the $\tau$ 
(four-momentum, spin state) in a format which is understood by 
{\tt TAUOLA FORTRAN}, and as a post processing step to return (insert) $\tau$ 
decay products to the primary event record.
Finally, role of such interfacing code is to calculate dedicated
weights from the production process information as well as from the decay, 
and unweight accordingly to standard MC procedures. Spin effects, 
electroweak corrections and also effects of anomalous couplings can be 
introduced in this way.

A rather modest version of {\tt TAUOLA Universal Interface} based on 
{\tt FORTRAN HEPEVT}
event record 
is described in \cite{Golonka:2003xt}, we will call it 
{\tt TAUOLA Fortran Interface}. A new version of {\tt TAUOLA Universal Interface}
  based on {\tt HepMC} \cite{Dobbs:2001ck}, the most popular  event record 
of C++, will be documented here. It also includes new 
functionalities.
We will call it {\tt TAUOLA C++ Universal Interface}, or, if no ambiguity 
could arise simply  {\tt TAUOLA C++ Interface} or  just
{\tt TAUOLA   Interface} if it is clear that the C++ not the
 FORTRAN version is in mind\footnote{The main class of 
{\tt TAUOLA C++ Universal Interface} is called {\tt Tauola}.}.
The {\tt PHOTOS}  generator for
QED bremsstrahlung in decays, which was previously distributed together with {\tt TAUOLA FORTRAN}, is not discussed in the present 
paper\footnote{{\tt PHOTOS} was distributed together with 
{\tt TAUOLA} since ref. \cite{Golonka:2003xt}. That is why it is present in our 
{\tt TAUOLA FORTRAN} but will be not used here.  }. It is now embodied  in a separate module
\cite{Barberio:1990ms,Barberio:1994qi} of the Monte Carlo simulation chain,
as the {\tt PHOTOS} generator has  found significant applications outside the domain of $\tau$ decays. 
That is why, in the future, we plan to distribute the C++ version of {\tt PHOTOS} 
separately \cite{photosC++}.

Important physics improvements, with respect to the {\tt TAUOLA FORTRAN Interface} implementation,
described in \cite{Golonka:2003xt}, comes with the code presented here. In
particular, transverse spin correlations have been implemented for
processes mediated by $Z/\gamma^*$ and genuine electroweak corrections
are now available for such processes.
With the new interface, it is rather straightforward to implement effects  beyond 
standard
model physics. Only  read-in data-tables should be replaced,
 no modification
to the code itself is needed. Further minor extensions 
include an algorithm to 
decay a single  $\tau$ with a user defined polarization, or the availability of  methods to
access generated $\tau$ leptons helicity states.
\section{Requirements  for {\tt TAUOLA Interface} }

\subsection{General Requirements}

For a $\tau$ decay to be generated it is enough to know its spin 
state and define the frame in which the decay should be performed. In case there are
more than one $\tau$ lepton in the final state, the quantum spin state of
both (or more) $\tau$ leptons must be provided in the form of a density matrix.
The exact  algorithm for the generation of spin correlations exist 
since the papers \cite{Jadach:1990mz,Jezabek:1991qp,Jadach:1993hs}. 
However, for the algorithm functioning, the density matrix must be known exactly as well.

In practice, the actual form of the spin density matrix 
(in our present paper it will be called $R_{ij}$, exactly as in the original 
{\tt TAUOLA FORTRAN} and its documentation) is often available  with some
approximations only.
With the increasing precision  of experiments,  
one may need to remove certain approximations introduced into $R_{ij}$.
Already now, our program  features, as an option,
complete spin effects in decays of $\tau$ pairs originating from the annihilation 
of quarks. Effects of genuine weak corrections are included, and an extension 
for the implementation of new physics signatures is straightforward. 
 
One should not forget that the density matrix itself is not the only place
approximations occur.  Effects of higher 
order QCD 
corrections need to be taken into account to define the kinematical configuration 
of initial partons 
used in spin density calculations (otherwise the density matrix for each individual 
process would have to be provided). At present, this is available in the 
 leading (collinear) approximation only.

Before we will discuss details of  specific implementation, let us recall 
first, the minimal list of steps the interface has to perform, independent 
of the programming language and data structure used.

\begin{enumerate}
\item The $\tau$ lepton or lepton pair(s) have to be localized in the 
event record. For processes mediated by $W$ bosons (or charged Higgs) 
 $\nu_\tau$ has to be localized as well.
\item
 If possible, the hard process leading to $\tau$ production has to be 
determined. This is necessary to control  transverse spin correlations.
Preferably minimal information from the host 
generators should be used. This is to reduce dependence on the host program\footnote{We 
target also applications when  event records will be filled by 
measured data. For example, 
measured $\mu^+ \mu^-$ events can be modified and final state muons replaced 
appropriately with $\tau^+ \tau^-$ pairs. Then generation of $\tau$ lepton decay is necessary.}.
\item Flavours and orientation of fields entering the production vertex for intermediate states, such as 
$Z/\gamma^*$, 
 have to be reconstructed too. This orientation (with respect to $\tau^\pm$ 
rest-frames) is necessary for calculation of the $\tau^\pm$ spin density matrix,
if spin of intermediate state is different from zero. 
\item The relative orientation of  $\tau^+$ and $\tau^-$ rest-frames 
should be established and respected by Lorentz transformations. 
\item Transformation of $\tau$ decay products from the $\tau$ rest-frame to 
lab frame has to be performed and the event record has to be completed  with $\tau$ decay products.
\end{enumerate}

\subsection{C++ Specific Requirements}

The C++ version of the {\tt TAUOLA Interface} implements all functionalities
of its predecessor, the {\tt TAUOLA  Interface} coded in FORTRAN \cite{Golonka:2003xt}.
It can be attached to any Monte-Carlo program where $\tau$'s
are generated, provided its output is available through a {\tt HepMC} \cite{Dobbs:2001ck} event record.

This condition is not very restrictive, is seems that {\tt HepMC} will
remain a generally accepted standard for the near future. However,
already now several different options for how {\tt HepMC} is used are
widespread. Possibility of the flexible  adaptation of our event record 
interface to different
options has been considered in the design,  drawing experience
from {\tt MC-TESTER} \cite{Golonka:2002rz,Davidson:2008ma}.
We have also
envisaged the possibility that {\tt HepMC} may one day be replaced by another
standard of event record, and we have provided an easy way to extend
the interface to a possible new event record standard.

\subsection{Object Oriented Event Records  -- The Case of {\tt HepMC}}
In adopting {\tt TAUOLA Interface} to the C++ event record format
 the difference between the {\tt HEPEVT} event record used in the FORTRAN
version of {\tt TAUOLA Interface} and  {\tt HepMC} event record  
which is used for the C++ based interface
has to be taken into account. 
In the first case the whole event was represented
by a common block containing a list of particles with their properties and
with integer 
variables denoting pointers to their origins and descendants.
 The {\tt HepMC} event structure
is built from vertices, each of them having pointers to their origins and descendants. Links between vertices represent particles or fields.
 In both, {\tt FORTRAN} and C++  
cases, the event is structured as 
a tree\footnote{At least in principle, because in practice its properties
may be rather of the graph of not universally defined properties.
This makes our task challenging.}, the necessary algorithms are  analogous, 
but nonetheless different.

In {\tt HepMC version 2.04}, an  event is represented by a {\tt GenEvent} object,
which contains all information regarding itself, including event id,
units used for dimensional quantities in the event and the list of produced particles. The particles
themselves are grouped into {\tt GenVertex} objects allowing access to mother
and daughter particles of a single decay. Vertices provide an easy way
to point to the whole branch in a decay tree that needs to be accessed,
modified or deleted if needed. The information of a particle  itself is stored
in a {\tt GenParticle} object containing the particle id, status and momentum
as well as information needed to locate its position in the decay tree.
This approach allows traversing the event record structure in several different
ways.

The {\tt HepMC} event record format is  evolving with time, making it necessary
 to adapt
the code to the new versions.
{\tt HepMC} version 2.05 is used as a reference. In the case of version 2.03 restrictions on methods for 
units
conversion have to be taken into account, for details see Appendix~\ref{subs:Issues}.
One should keep in mind that future adaptations to {\tt HepMC} changes may restrain
backward compatibility.

Evolution of the {\tt HepMC} format itself is not a crucial problem.
In contrary, conventions how physics information is  filled into {\tt HepMC}
 represent the source of main technical and also physics 
challenge for our interface. 
This is quite similar to the previous
{\tt HEPEVT - FORTRAN} case. Let us discuss this point in more detail now.

\subsubsection{Event Record Structure Scenarios}
While many Monte-Carlo generators (eg. {\tt PYTHIA 8.1} \cite{Sjostrand:2007gs}, 
HERWIG++ \cite{Bahr:2008pv})
store events in {\tt HepMC} format, the  representations of
these events are not subject to strict standards,  which can therefore
vary between Monte-Carlo generators or even physics processes. Some examples
of these variations include the conventions of status codes, the  way
documentary information on the event is added, the direction of pointers at a vertex
and the conservation (or lack of conservation) of energy-momentum at a vertex.
Below is a list of scenarios we have observed in Monte-Carlo
generators used for testing the code.

This list will serve as a declaration for convention of  {\tt HepMC} filling, which  the 
interface should  be able to interpret correctly.

\begin{itemize}
  \item \textbf{4-momentum conservation} is assumed for all vertices in the event record.
  \item \textbf{Status codes:} only information whether given particle is incoming, outgoing or intermediate will be used,
  \item \textbf{Pointers at a vertex} are assumed to be bi-directional. That is, it is possible to traverse the record structure from mother to daughter and from daughter to mother along the same path.
\end{itemize}

\noindent
\textbf{ Extensions/Exceptions} to this specifications  are handled in some cases. We will call them
options for conventions of event record filling.
  \begin{itemize} 
    \item  Vertices like $\tau^\pm \rightarrow \tau^\pm$ and $\tau^\mp \rightarrow \tau^\mp \gamma$ 
           where 4-momentum conservation is not preserved, but this
           non-conservation is balanced, for example, between the two branches
outgoing from $Z$. 
    \item  Lines representing
           intermediate bosons may be missing. In fact this may be unavoidable, if several 
           diagrams contribute simultaneously. In that case, our algorithm makes a choice
           based on an approximation that
           only the dominant  single diagram is considered
           and an intermediate boson state is defined accordingly on the fly. Other possible treatments:
           statistical choice of the dominant process, or calculations based on higher order
           matrix elements for the hard process, are not available at present.
    \item
           As in the FORTRAN cases, we expect that  new  types of 
           conventions for filling the event record
           will appear, because of physics motivated requirements.
           Unfortunately, the resulting options do not always guarantee
           an algebraically closed structure.  
           Host program specific patches  may need to be defined for the 
           {\tt TAUOLA interface}. 
           Debugging it could be time consuming, and will need to be repeated for every new
           case.
   \end{itemize}

Detailed
conventions for the actual filling of physics information into {\tt HepMC} format
is defined by authors of each Monte Carlo program.
 In future,  an important special case of event records filling with
information extracted from experimentally observed event (eg. $Z\to \mu^+\mu^-$
 modified later to $Z\to \tau^+\tau^-$) should be allowed.
  Obviously, a new type (or types) of {\tt HepMC} filling will then appear.

\section{Design}

The structure of our code is documented using Doxygen standards \cite{Doxygen} and is
presently available from the project web page \cite{tauolaC++}. The
source code for this web page is also available in our package
distribution.  Doxygen documentation can be thus compiled on a users
platform, and hence provide documentation which matches the actual version of the
distribution. 

Let us present here briefly the directory structure and list the main classes
with a short description of their functionality.

\subsection{Interface Structure and Responsibilities}

\begin{figure}[h!]
\centering
\includegraphics[scale=0.6]{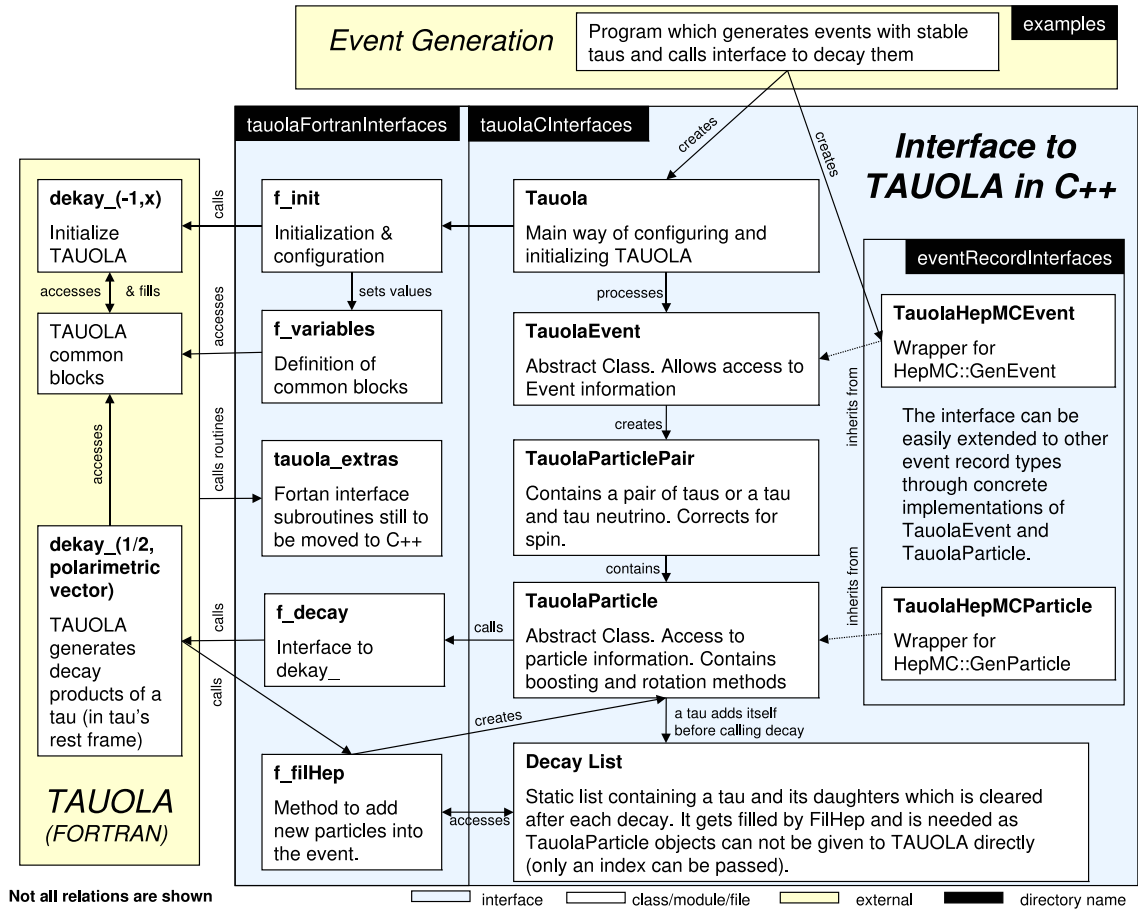}
\label{fig:design}
\caption{{\tt TAUOLA C++ Interface} class relation diagram}
\end{figure}

The choice of splitting the source code into three main modules, 
see Fig.~\ref{fig:design} (blue part),
allows to separate the FORTRAN related code from the abstract C++ interface
and the concrete implementation of the interface created for the appropriate
event record.

\begin{itemize}
  \item {\bf Tauola Fortran Interface}\\
       This part of the code provides an interface 
       to the FORTRAN library of {\tt TAUOLA}. In particular,  routines
       necessary for library initialization and wrapper for routine invoking the decay of a single 
       $\tau$. Parts of the interface code are still left in FORTRAN, but  can be rather easily
       rewritten to C++. 
       The most important method, {\tt filhep\_},
       is already implemented  in C++. Its FORTRAN predecessor writes single particles to
       the {\tt HEPEVT} common block. At present the method {\tt filhep\_} inserts the particle 
       into the {\tt HepMC} event record but remains to be called from the FORTRAN library.
       For further details see Appendix \ref{Interface to TAUOLA}.
  \item {\bf Tauola C++ Interface} \\
       The abstract part of the interface to the event record.
       The class {\tt TauolaEvent} contains information regarding the whole event
       structure, while {\tt TauolaParticle} stores all information regarding a single particle.
       All particles used by the interface are located in the event in the form of
       a list of {\tt TauolaParticle} objects.
       The last class located here, {\tt TauolaParticlePair}, is the core of all polarization
       and decay algorithms. They are independent from the event record used by the interface
       as they operate on these two abstract classes presented above.
  \item {\bf Event Record Interface} \\
       The event record implementation classes. All classes stored here represent
       the implementation of specific event record interfaces and are responsible for reading,
       traversing and writing to the event record structure.
       Only {\tt TauolaEvent} and {\tt TauolaParticle} classes must be implemented.
       The {\tt HepMC} event record interface is implemented
       through {\tt TauolaHepMCEvent} and {\tt TauolaHepMCParticle}.
\end{itemize}

\subsection{Directory Structure}

\begin{itemize}
\item {\bf src/eventRecordInterfaces/ } - source code for classes which interface with {\tt HepMC}.\\
  Classes:
  \begin{itemize}
  \item { \bf TauolaHepMCEvent} - Interface to HepMC::GenEvent objects. 
  \item { \bf TauolaHepMCParticle} - Interface to HepMC::GenParticle objects. 
  \end{itemize}   

\item {\bf src/tauolaCInterfaces/ } - source code for general {\tt TAUOLA Interface} classes, such as those
  responsible for spin correlations and boosting.  \\
  Classes:
  \begin{itemize}
  \item { \bf DecayList } - Storage class for keeping track of TauolaParticles and their indices 
  \item { \bf Tauola } - Controls the configuration and initialization of {\tt TAUOLA FORTRAN}.
  \item { \bf TauolaEvent } - Abstract base class for event information.
  \item { \bf TauolaParticle } - Abstract base class for particles in the event. This class also handles particle boosting. 
  \item { \bf TauolaParticlePair } - Contains two objects of type TauolaParticle that are related by the same mother. Spin correlations and other minor algorithms are handled here. 
  \end{itemize}

\item {\bf src/tauolaFortranInterfaces/ } -  interface to {\tt TAUOLA FORTRAN} routines and common blocks. \\
  Files:
    \begin{itemize}
    \item { \bf f\_Decay} - contains a wrapper for the {\tt TAUOLA FORTRAN} routine for decaying 
$\tau$'s (DEKAY).
    \item { \bf f\_FilHep } - provides a method which {\tt TAUOLA FORTRAN} calls to fill a $\tau$ decay product into the
      event record.
    \item { \bf f\_Init } - contains a wrapper for the {\tt TAUOLA FORTRAN} routines for tauola initialization.
    \item { \bf f\_Variables } - contains definitions of {\tt TAUOLA FORTRAN} routines and common blocks used by other methods in tauolaFortranInterfaces. 
    \item { \bf tauola\_extras.f } - contains extra FORTRAN routines (taken from the {\tt TAUOLA  Interface} in {\tt FORTRAN}) which should ultimately be migrated to C++.
    \end{itemize}

\item {\bf src/utilities/ } - source code for utilities that help in debugging and plotting distributions.\\
  Classes:
  \begin{itemize}
  \item { \bf Log} - General purpose logging class that allows filtering out output messages 
      of {\tt TAUOLA C++ Interface} and keeps statistics regarding tauola run.
  \item { \bf Plot} - Simple class that gathers data for some useful debug plots.
  \end{itemize}   

  \item {\bf examples/ } - examples of different {\tt TAUOLA C++ Interface} uses
    \begin{itemize}
    \item {\bf taumain\_stand\_alone\_example} - stand alone example with a simple 
      $e^+e^- \rightarrow \tau^+\tau^-$ event in  {\tt HepMC} format
     and then $\tau$'s decayed by {\tt TAUOLA}.
    \item {\bf single\_tau\_gun\_example } - example of {\tt TAUOLA} linked with pythia 8.1 and used to decay single $\tau$ selected from the event record.
    \item {\bf taumain\_pythia\_example } - example of  {\tt TAUOLA} linked with pythia 8.1, and decay chain  analysed with  {\tt MC-TESTER}.
    \end{itemize}   
  \item {\bf SANC/} - code for the computation of  electroweak corrections.
  \item {\bf include/} - directory for the header files.
  \item {\bf lib/ } - directory for the compiled  libraries. 
  \item {\bf documentation/ } - contains doxygen documentation and this latex file.
  \item {\bf tauola-fortran/ } standard  {\tt TAUOLA FORTRAN} distribution 
exactly as described in \cite{Golonka:2003xt}. It is kept intact and
 is prepared for future updates, see
ref.~\cite{Actis:2009gg} for details of that project.
\end{itemize}

\subsection{Algorithm Outline}
\label{sect:Outline}

An overview of the algorithm for  the {\tt TAUOLA Universal Interface} is
given below,  for more detail the reader
should refer to the project's Doxygen documentation ~\cite{tauolaC++}.
Documentation of the  {\tt TAUOLA FORTRAN Interface}
~\cite{Golonka:2003xt} describes some aspects of the spin
correlation algorithm which are also relevant to this interface.

The first step is creation of a {\tt TauolaHepMCEvent} object from
a {\tt HepMC::GenEvent} event record. At this step the units for dimensional
quantities (four-momenta, masses, etc.) are checked, and if needed the {\tt HepMC} event record is
reset to use GEV and MM ensuring proper execution of $\tau$ decay library.
 After a {\tt TauolaHepMCEvent} is created the
{\tt decayTaus()} method should be executed by the user's code\footnote{Prior to this step
the user may want to execute {\tt Tauola::decayOne(...)} for $\tau$ leptons, where {\tt TAUOLA Universal Interface} is 
expected not to work properly. For details see Appendix~\ref{TAUOLAgun}.
},
invoking the following process:

\begin{enumerate}
\item The {\tt HepMC} event record is traversed and a list  of all stable $\tau$'s in the event is created.
\item From each found $\tau$ location, the tree is traversed backwards so
  that information about the production process can be extracted
  and used for the calculation of the spin density matrix.
\item The siblings of the $\tau$ are identified through common
  parents,  i.e. requiring that they are produced at the same {\tt HepMC} vertex. In
  cases such as $\tau \rightarrow \gamma \tau$, the parent(s) are defined
  as the particle(s) which produced the first $\tau$; 
  $\tau$ and $\nu_\tau$ siblings are paired to the $\tau$.
\item The density matrix is set-up using information about the
  $\tau$-pair and their parent type (for $Z/\gamma$ processes,
  grandparent information is also required). This is described in
  detail in Sec. \ref{dwaTrzy}. The density matrix assumes
  a center-of-mass frame for the $\tau$-pair, with the $\tau$'s and their
  grandparents orientated as shown in Figure \ref{fig:designKB}.

\item The pair is then decayed by executing the {\tt DEKAY} routine 
      for each $\tau$ in the pair. The {\tt DEKAY} routine
is stored in  the {\tt tauola-fortran} directory, for details see Appendix
\ref{subsubsec:routines}.

\item A spin weight is calculated using the polarimetric vectors
  returned from {\tt TAUOLA FORTRAN} and the density matrix previously set-up (described in
  Sec. \ref{dwaTrzy}).
\item If the decays are rejected, the pair is decayed anew and the
  process is repeated until the decays are accepted. In this way unweighting of spin 
  effects is performed.

\item Once accepted, the decay products are added into the event record with the procedure
      as follows:

  \begin{enumerate}
  
\item As the density matrix is only valid in the special reference
  frame of Figure \ref{fig:designKB}, the $\tau$-pair are boosted and rotated into this hard process frame.

\item The DEKAY routine of {\tt TAUOLA FORTRAN} is executed with state = 11 or 12
  (write). This initiates {\tt TAUOLA FORTRAN} to return the daughter
  information via the {\tt filhep\_} routine (see Section \ref{subsubsec:routines}).

\item The $\tau$'s status code is changed from ``1'' (stable particle)
  to ``2'' (intermediate particle).

\item A new object HepMC::GenParticle is created for each daughter and the
  appropriate tree structure is created and added into the event.

\item Each daughter is boosted using the $\tau$'s 4-momentum
  (as {\tt TAUOLA} constructs a decay for a $\tau$ at rest) to the hard process frame. 

\item The $\tau$'s and their decay products are boosted back into the
  laboratory frame.

\end{enumerate}

\item As the final step, the position of vertices containing the $\tau$'s and their decay products
      is set according to the $\tau$'s momentum and lifetime.

\end{enumerate}

The underlying HepMC::GenEvent is hence modified with insertion of  $\tau$ decay products.
All that remains is to convert the event back to its initial units, which is done
via the {\tt eventEndgame() } routine of the {\tt TauolaHepMCEvent} class.

\section{Calculation of Spin Correlations}
\label{dwaTrzy}

If more than one $\tau$ lepton is present in a final state, then not only is the individual
spin state for each $\tau$ 
necessary for  proper generation, but the complete correlation matrix
of all $\tau$ leptons must be taken into account  as well. In the case 
of $\tau$-pair production,  the standard algorithm explained 
in \cite{Jadach:1990mz,Jadach:1993hs}
can be used  without much modification. 
For the single $\tau$ produced in a $\tau,\;\nu_\tau$ pair,
it is convenient to use the same algorithm as well, 
even though it is not necessary from the physics point of view.

Let us describe now the algorithm given in  
Refs.~\cite{Jadach:1990mz,Jadach:1993hs}. We will
use definitions and notations from that papers as well.
Spin correlations and spin polarization effects can be  simulated by accepting or rejecting 
a pair of generated $\tau$ decays based on a weighting factor {\tt wt}.

\begin{equation}
wt=\displaystyle \frac{1}{4} \sum_{i,j=0}^{4} h^1_i h^2_j R_{ij} 
\label{weight}
\end{equation}
where $h^1$ and $h^2$ are the polarimetric vectors
for the $\tau^+$ and $\tau^-$ respectively and $R_{ij}$ is the 
density matrix associated with the $\tau$ production vertex. 
 The matrix  $R_{ij}$ depends on the mechanism 
and particular kinematical configuration of
$\tau$ pair production. The  $h^1_i$, $h^2_j$ depend on the respective 
decays of $\tau^+$ and $\tau^-$.  
The solution can be used for $\tau-\nu_{\tau}$ production as well.
In this case $\nu_{\tau}$ decay is not performed and its polarimetric vector is set to $h=(2,0,0,0)$.

A pair of $\tau$ decays should be accepted if the weight is greater than
a randomly generated number between 0 and 1.
If this condition fails, the $\tau$ pair decays should either be rejected and
regenerated, or rotated\footnote{Rotation instead of rejection increases 
efficiency by a factor of 4. This however only affects the generation of $\tau$ 
lepton decays and represents a small fraction of the total time of
constructing the event.},
and the weight recalculated. The production process does not need to be 
reprocessed.

The density matrices, $R_{ij}$, for the most standard
processes of $\tau$-pair production, are documented below.
The following frame 
convention\footnote{Fig.~\ref{fig:designKB} illustrates our choice too.
There however the reaction frame is rotated by angle   $\theta$ around x axis.} 
was adopted:
\begin{itemize}
  \item The $\tau$-pair's center of mass system  is used 
  \item The $\tau^{+}$ (if present) lies along the positive z axis
  \item The $\tau^{-}$ (if present) lies along the negative z axis
  \item The incoming beams (if present) lie in the z-y plane.
  \item If applicable, the incoming antiquark (or antilepton) y momentum component is positive.
\end{itemize}

$h$ is defined such that $h_0$=1 and $h_{1,2,3}$ form the polarimetric vertex
returned from {\tt TAUOLA FORTRAN} (see DEKAY in Appendix \ref{subsubsec:routines}).
 The $h$  is defined in the rest frame of the $\tau$ it 
was calculated for. One should stress that formally speaking  $R_{ij}$
does not represent a Lorentz invariant object. Its first index is defined in the
rest frame of the $\tau^+$, whereas the second index 
is in the frame of the $\tau^-$.

In the following subsections we will list the form of $R_{ij}$ for the most 
commonly used processes of $\tau$-pair (or $\tau$, $\nu$) production.

\subsection{Form of $R_{ij}$ for Standard Processes}
\subsubsection{$Z/\gamma \rightarrow \tau^+ \tau^-$}
\label{subsubsection:z/gamma}

\[R=\begin{pmatrix} 
1 & 0 & 0 & 2P_z(cos\theta)-1 \\
0 & 0 & 0 & 0 \\
0 & 0 & 0 & 0 \\
2P_z(cos\theta)-1 & 0 & 0 & 1 
\end{pmatrix}\]

where $P_z$ is calculated from the square of the matrix elements of the 
Born-level $2 \rightarrow 2$ process $f \bar{f} \rightarrow \tau^+ \tau^-$.
\[P_z(s,\theta) = \frac{\frac{d\sigma(s,\theta,+,+)}{d\Omega}}{\frac{d\sigma(s,\theta,+,+)}{d\Omega}+\frac{d\sigma(s,\theta,-,-)}{d\Omega}} \]
$\theta$ is the angle between the incoming antiparticle beam and outgoing $\tau^+$. If the incoming beam 
cannot be reconstructed from the event record the average of $P_z$ should be used 
(which is equal to $P_z(cos\theta=0)$). The ``$+$'' denotes that 
the spin states of $\tau^+$ and $\tau^-$ are parallel to the $\tau^+$ momentum.
For ``$-$'' it is placed in the opposite direction.

The spin correlation matrix explained above is approximate and is valid only for 
longitudinal spin effects. The object $R_{i,j}$ is nonetheless prepared to 
host complete spin effects. For that purpose information available  from the module based on {\tt SANC} 
can be used, see Appendix \ref{section:SANC}.
Further advantage is that  genuine weak effects can be calculated and 
included as a weight\footnote{In a similar way one can implement effects of new physics, 
such as Z'
into the program. With the help of our interface effects of weak corrections 
on the cross section, and not only on polarization can be installed with additional 
 weights.}  not only on polarization but on the cross section as well.
This was found to be numerically important \cite{Adam:2008pc,Adam:2008ge} for final states
of virtuality largely surpassing the $Z$ mass and should be taken into account prior to the implementation of  new physics effects.

In the formulas above, the hard process kinematical variables  $s$ and $\theta$ have to be known
for each event. Those variables, together with the flavour of the incoming beam are used by a module for calculating 
electroweak corrections or the function $P_z$.\footnote{
 The principle behind our 
solution, is quite similar  to the one used in {\tt PHOTOS} Monte Carlo
where it is was shown \cite{Golonka:2006tw} to be  valid up to NLO (QED) precision level. 
It relies on factorization properties of fully differential distributions into the appropriately chosen
Born level ones and emission factors.
To achieve such precision in the case of spin correlations in proton-proton
collisions, rather challenging  work on QCD matrix elements would be  necessary.
At present,
our predictions will not be better than LL {\it on spin effects}. It is known \cite{Kleiss:1990jv},
that the solution can not be constructed beyond NLO.}

To apply the method we need to identify the four momenta of the $\tau^+$ and $\tau^-$ pair first. In the rest frame of the pair the two effective partons leading to the hard process are not necessarily back to back. Two scattering angles $\theta_1$ and $\theta_2$ can be thus reconstructed.  The angle $\theta_1$ is between the $\tau^+$ and the first incoming state,
$\theta_2$ is between $\tau^-$ and the second one\footnote{We choose the first incoming state to be antiparticle.
 }. Both angles are 
calculated in the rest frame of the $\tau$ pair. 
The average angle $\theta^\bullet$ accordingly to the description given in \cite{Was:1989ce} is taken:
$\cos\theta^\bullet=\frac{\sin\theta_1\cos\theta_2+\sin\theta_2\cos\theta_1}{\sin\theta_1+\sin\theta_2}$.

If events originate from  generator such as {\tt PYTHIA}, 
the flavour of incoming partons is explicitly given or it 
can be calculated using information encoded in the event record. 
In the generic case this information is not available, and one will have to rely on  measured structure 
functions and statistical choice. This is the method, for example, to be applied in the analysis where one
reconstructs
   from experimental  data decays $Z\to \mu^+ \mu^-$  (or $W \to \mu \nu_\mu$)
 and then replaces $\mu$ by the
Monte Carlo generated
   $\tau$ object.
The resulting 
uncertainty on one hand will not be large; angular polarization dependence of $d$ and $u$ quarks
couplings to a $Z$ is not that different. On the other hand, a mismatch between choosing quark
and antiquark may have larger effect.


The density matrix presented above features only longitudinal spin correlations. 
Once the matrix $R$ is replaced with the one featuring complete spin effects,
 see Appendix \ref{section:SANC}, this is 
passed to  the results of simulation as well, without any need of further changes.

\subsubsection{$H^0 \rightarrow \tau^+ \tau^-$ and $A^0 \rightarrow \tau^+ \tau^-$ and mixed $A^0/H^0 \rightarrow \tau^+ \tau^-$}
The complete density matrix for a scalar neutral Higgs boson $H^0$ is rather simple,
\[R=\begin{pmatrix} 
1 & 0 & 0 & 0 \\
0 & 1 & 0 & 0 \\
0 & 0 & 1 & 0 \\
0 & 0 & 0 & -1 
\end{pmatrix}\]

it is also true for a pseudoscalar neutral Higgs boson $A^0$

\[R=\begin{pmatrix} 
1 & 0 & 0 & 0 \\
0 & -1 & 0 & 0 \\
0 & 0 & -1 & 0 \\
0 & 0 & 0 & -1 
\end{pmatrix}\]

A mixed $A^0/H^0 \rightarrow \tau^+ \tau^-$   represents only a slightly more complicated case:
\[R=\begin{pmatrix} 
1 & 0 & 0 & 0 \\
0 & \frac{(\beta cos\phi)^2 - sin\phi^2 }{(\beta cos\phi)^2+sin\phi^2} & -\frac{2\beta cos\phi sin\phi}{(\beta cos\phi)^2+sin\phi^2} & 0 \\
0 & \frac{2\beta cos\phi sin\phi}{(\beta cos\phi)^2+sin\phi^2} & \frac{(\beta cos\phi)^2 - sin\phi^2 }{(\beta cos\phi)^2+sin\phi^2} & 0 \\
0 & 0 & 0 & -1 
\end{pmatrix}\]
where $\beta = \sqrt{1-(\frac{2M_{\tau}}{M_{A^0/H^0}})^2}$
and $\phi$ is the scalar-pseudoscalar mixing angle.

\subsubsection{$W^{\pm} \rightarrow \tau^{\pm} \nu$}
\label{subsubsection:W}

 For $W$ the matrix $R_{ij}$ takes the following form:

\[R=\begin{pmatrix} 
1 & 0 & 0 & 1 \\
0 & 0 & 0 & 0 \\
0 & 0 & 0 & 0 \\
1 & 0 & 0 & 0 
\end{pmatrix}\]

\subsubsection{$H^{\pm} \rightarrow \tau^{\pm} \nu$}
For charged Higgs decay the matrix $R_{ij}$ differs from the $W$ case by signs only:
\[R=\begin{pmatrix} 
1 & 0 & 0 & -1 \\
0 & 0 & 0 & 0 \\
0 & 0 & 0 & 0 \\
-1 & 0 & 0 & 0 
\end{pmatrix}\]

\subsection{Cases of Partly defined Hard Processes}
\subsubsection{$\tau^+ \tau^-$ Pair with Multiple Parents and Sisters}
If a $\tau^+ \tau^-$ pair is found with multiple parents rather
than a single parent, the parent type is assumed to be a
$Z/\gamma$ with the 4-momentum  reconstructed from
the 4-momentum of the $\tau$ pair. The density matrix  
\ref{subsubsection:z/gamma} is used. For more details on the construction
of effective incoming beams see  Section~\ref{dwaCztery}.

\subsubsection{$\tau \nu_{\tau}$ Pair with Multiple Parents and Sisters}
If a $\tau^+ \nu$ pair is found with multiple parents rather
than a single parent, the parent type is assumed to be a
$W^{\pm}$ with the 4-momentum  reconstructed from 
the 4-momentum of the $\tau\; \nu_\tau$ pair. The density matrix 
\ref{subsubsection:W} is used. 

\subsubsection{Single $\tau$ and Multiple Unpaired $\tau$'s}
By default a single $\tau$, with no $\tau$ or $\nu_{\tau}$ pairing, is  treated
as unpolarized and spin effects are ignored. However, in such a case the {\tt TAUOLA decayOne} method,
see Appendix~\ref{TAUOLAgun}, can be used if including spin effects is requested. This method can be applied
by the user, in such a case the spin state of a single $\tau$ can be imposed. The method can be used even for 
multiple $\tau$ final state and exact spin correlations.
In those cases a user defined quantization frame for each  $\tau$ may be necessary. For each $\tau$ a distinct routine for boosting from its rest-frame to the lab-frame
may be necessary. The appropriate method is explained in the Appendix mentioned above.

Another user defined option (which may become in the future a default part of the {\tt TAUOLA Interface}) is when pairing can not be done on the basis of inspecting $\tau$
mother(s) but
can be performed
 according to the closeness of the reconstructed invariant
masses of the pairs
to the masses of $W's$ or $Z's$ (if the appropriate Standard Model processes are under considerations). In general, as such configurations will often 
appear  for processes of new physics, again a user defined and hard process dependent solutions based on the {\tt TAUOLA decayOne} method,
might be the only option.

\subsection{Quantum Entanglement and Helicity States}

One of the convenient feature of Monte Carlo simulation is the availability of 
variables used in the generation of hard processes. However, it is often not possible to define
in an exact manner what is the energy transfer, eg. in the Z propagator, if several
diagrams contribute simultaneously. Nonetheless use of such
information is tempting as it would be helpful to validate algorithms used in experimental analysis
for defining observables aimed
at the measurement of the Z lineshape.

Another example of such useful variables  are  the helicities  of  $\tau^+$ and $\tau^-$.
Even though,  it is possible to attribute such variables only in the ultrarelativistic limit
and its use can be restricted to a downgraded physics approximation only (quantum entanglement
\cite{Einstin:1935rr} ignored), helicity  was offering invaluable help at LEP time,
 for the measurement of $\tau$ polarization \cite{Heister:2001uh}. 

Our program provides helicity states of $\tau^+$ and $\tau^-$ even in cases 
when exact spin effects are taken into account in the generation, as is the case of processes mediated by 
intermediate  $Z/\gamma^*$ state or in Higgs boson decay. 
We just attribute helicity states for the $\tau$ lepton after decays
are already generated and accepted.
The approximation used in calculation of these helicities is explained in Section~\ref{Exact} 
and technical aspects of our solution
are given in Appendix~\ref{subsection:eHelicity-ew}. This information
can be used for solutions similar to the ones in  \cite{Heister:2001uh},
but this time for LHC purposes

\subsection{Handling of  Events with Bremsstrahlung or Parton Shower Activity}
\label{dwaCztery}
Obviously, there are cases, when spin correlations calculated from  Born level 
processes can not be applied directly. Good examples are:
(a) $Z \to \tau^+ \tau^- \gamma$,
(b) $f1 + f2 \to  Z + X$, $Z \to \tau^+ \tau^-$ where the intermediate
state $Z$ is explicitly stored,
 (c) $f1 + f2  \to \tau^+ \tau^- X$  or $f1 + f2  \to \tau^+ \nu_\tau X$  where the $Z$ state is not 
available. Here $X$ represents parton shower and/or final state bremsstrahlung.
The fist step, necessary eg. for calculation of the spin correlation 
matrix, is to reconstruct effective Born level variables $s$ and $\theta$ and the incoming state flavors; the arguments of
the function $P_Z$. This is equivalent to the construction of  effective incoming and outgoing $\tau$ fields.
Let us discuss now our cases:

\begin{itemize}
\item[a)]
 An additional photon is  added to the $\tau$ with which it forms
 smaller virtuality. For construction of the transformation between the laboratory frame and rest frame 
of $\tau$'s the effective state $Z-\gamma$ frame is used instead of intermediate $Z$.
This choice is motivated by inspection of properties of spin amplitudes.
In such a frame the effective  incoming states f1 f2 will not necessarily be back to back. The average of the two directions ($\theta^*$, $\theta^\bullet$ as explained earlier in this subsection)
can be used. The virtuality of the $Z$ is nonetheless used in the effective Born calculation.

\item[b)] Additional fields $X$ representing parton showers should be subtracted from $f1$ or $f2$, 
preferably from the one with which it forms smaller virtuality.
This is motivated by inspection of the spin amplitudes.
 Once the effective incoming states are constructed, the definition of 
boosting routines is straightforward. 

\item[c)] This case  is a combination of the above  two. Additional fields should be subtracted from  f1 f2 or considered
as originating from the intermediate $Z$ reconstructed on flight together with
$\tau^+$, $\tau^-$. The minimalization of virtuality should be used as a guide whenever combinatorial
choice has to be made. However electromagnetic
charge or colour charge should not be neglected. Obviously photons should not be combined with neutrinos nor gluons with leptons.
\end{itemize}

The approach presented above  is explained in ref \cite{Golonka:2003xt} in more detail. The principle 
is based on the simplest factorization properties of SM/QCD matrix elements. 
In particular the assumption is made that   photon(s) can be treated as (nearly) collinear with one of the final state 
$\tau$ leptons. Then the kinematics can be built on the 
$\tau^+ \tau^-$ pair rest frame and the effective  incoming states. The 
 z axis is taken along effective  
 antiparticle incoming state and y axis is of  half plane  including $\tau^-$. 
The second effective beam is  placed then in zy plane as well, even though it is not back to back
with the first one.
 For calculation of $R_{ij}$ the  variable $s$ should be calculated from effective  incoming states.
The scattering angle can be calculated in the $Z$ frame using formula for $\theta^*$ or $\theta^\bullet$, see ref. \cite{Was:1989ce}. 
In the collinear limit for photon emissions,  Lorentz transformation between the $\tau^+$, $\tau^-$ and $Z$ rest
frames is reduced to a simple  boost along the $\tau^+-\tau^-$ flight directions in the $Z$ rest-frame.

Further improvements with respect to that description
require explicit use of higher order matrix elements. The described  approximation is already quite good
and  works up to $\alpha_{QED} / \pi \simeq$ 0.1 - 0.2 \% precision level,
for observables where it is not requested explicitly that high $p_T$ photons are present.

Situation with initial state parton shower emissions is similar, however in this case
the omitted terms for $R$ calculation may be of the order of $\alpha_{QCD} / \pi$. Thus significantly larger, 
but still at the level of ten  percent or so.

If the event record under 
study is originating from experimental data, the flavours of the incoming partons
 can not be known from the event record itself. 
Instead, information in PDFs can be used to attribute such flavours on a
statistical basis and then used in the calculation of the matrix $R_{ij}$. 

\subsection{Exact Spin Effects and Helicity States}
\label{Exact}
Earlier in this Section  we have listed some examples of spin polarization, 
spin correlation and density matrices. Those matrices can be easily 
changed/replaced by the externally calculated ones
for the sake of studies on new physics phenomena, like  
a consequence of certain types of spin correlations on the signal/background
separation. In that  way, effects of 
new physics or ad hoc modification of spin effects component by component,
can be easily included.

For calculating complete spin correlations, including the effects 
of genuine weak corrections in the case of the single boson mechanism of $\tau$-pair production, 
another solution 
is also prepared, see  Section \ref{Sect:electroweak}.
It works in the  case of  an intermediate  $\gamma^*/Z$ state  
 produced in the annihilation of a pair of quarks.  This can serve as an example for other processes
which can be implemented in a similar manner.

Let us stress that in the case of including complete spin effects
 it is not possible to attribute helicity states to the produced $\tau$ leptons.
This can be done only  in an approximate way. For that purpose,
a modified formula (\ref{weight}) can be used.
The weight 
for each helicity configuration is:
\begin{equation}
weight(\;,\;)=\displaystyle \frac{1}{4} \sum_{i,j,i'j'=0}^{4} h^1_{i'} h^2_{j'} R_{ij} P^1_{ii'} P^2_{jj'}
\label{weights}
\end{equation}

The matrices $P^{1,2}$ (spin projection operators) read
\[P^{1,2}=\begin{pmatrix} 
1 & 0 & 0 & \pm 1 \\
0 & 0 & 0 & 0 \\
0 & 0 & 0 & 0 \\
\pm 1 & 0 & 0 & 1 
\end{pmatrix}\]

As a consequence for {\tt Hel1=}$\pm 1$ and {\tt Hel2=}$\pm 1$
\[\begin{pmatrix} 
weight(+,+)= & (h^1_{0}+ h^1_{3}) & (h^2_{0}+ h^2_{3}) & (R_{00}+R_{03}+R_{30}+R_{33}) \\
weight(+,-)= & (h^1_{0}+ h^1_{3}) & (h^2_{0}- h^2_{3}) & (R_{00}-R_{03}+R_{30}-R_{33}) \\
weight(-,+)= & (h^1_{0}- h^1_{3}) & (h^2_{0}+ h^2_{3}) & (R_{00}+R_{03}-R_{30}-R_{33}) \\
weight(-,-)= & (h^1_{0}- h^1_{3}) & (h^2_{0}- h^2_{3}) & (R_{00}-R_{03}-R_{30}+R_{33})
\end{pmatrix}\]
The actual helicities can be then attributed by unweighting the above helicity weight.

It is no omission that we are not explicit on  sign conventions 
for the generated helicity variables {\tt Hel1} and {\tt Hel2}.
 This depend on the particular choice of boosting routine. 
Our  choice adopted here is, that for the $\tau$ pair produced at the $Z$ 
peak on average both {\tt Hel1} and {\tt Hel2} will be negative and 
 {\tt Hel1 = Hel2}. 

\section{Electroweak Corrections and Refined Spin Effects}
\label{Sect:electroweak}
\subsection{External Calculation of the Spin Density Matrix $R_{ij}$}

Our program is equipped with methods for calculating simple spin density matrices 
for most of the interesting hard processes. These methods are explained in the text of the paper, 
see Section \ref{dwaTrzy}. 
In some cases, notably in the case of $\tau^+ \tau^-$ produced from the annihilation of a pair of quarks,
the standard density matrices may not be sufficient for some applications. A more exact solution
is also available.  Instead of a native $R_{ij}$ density matrix, an externally calculated one can be used. 

The solution is based on  {\tt SANC} library ~\cite{Andonov:2008ga,Andonov:2004hi} 
for calculation of electroweak corrections%
\footnote{It may serve as an example of how other 
calculations featuring heavy $Z'$ boson, for example, may be used in our interface.}.
With its help the density matrix $R_{ij}$ for $q \bar q \to \tau^+ \tau^-$ process
can be calculated as a function of the incoming state flavour and Born level variables  (Mandelstam
$s$ and  scattering  angle $\theta$).
Additional two  weights are also provided, which include
 the matrix elements squared and averaged over the spin. 
For additional weights genuine weak corrections are respectively switched on and off.
This may be helpful for the evaluation of genuine weak corrections for states
of large $s$, significantly above the $Z$ peak, where they become sizable.
See eg. refs. \cite{Adam:2008pc,Adam:2008ge}. 

For better modularity of the interface and to speed up execution of the program, pretabulation is used. 
At first, a dedicated module has to be invoked, as will be explained later. 
In such dedicated runs, $R_{ij}$ is calculated and stored in a lattice of 
 ($s$ and $\cos\theta$) points.

Later, in the actual execution of our interface, 
these pretabulated values of $R_{ij}$ are interpolated to the actual phase space point.
 For this purpose, the standard bilinear interpolation algorithm 
is used. Additionally, in order to avoid numerical errors, for $\cos\theta$ values near -1 and 1 we're using the linear extrapolation algorithm.

Pretabulation is prepared for  3 domains of  $s$: around the $Z$ peak, close to the $WW$ 
pair production threshold and over a broad energy range. The actual choice for 
pretabulation zones is  $ 85\; $GeV$ < \sqrt{s} < 110\; $GeV,  
$160\; $GeV$< \sqrt{s}< 220\; $GeV\footnote{ In the case when the application featuring  $Z'$ is used this
pretabulation zone should be replaced, for example by 
$ M_{Z'}- 3\Gamma_{Z'}< \sqrt{s} < M_{Z'}+ 3\Gamma_{Z'}$.},
$6\; $GeV$< \sqrt{s}< 17\; $TeV.  For $s$ below $36\; $GeV$^2$ the analytic form taken from 
ref. \cite{Jadach:1985ac} is used.
 It features all spin and mass effects, but electroweak corrections, and even 
$Z$ exchange, are not taken into account. This is reasonable for $s<36$ GeV$^2$
(up to, say, 100 GeV$^2$).

The advantage of this solution is that results of  {\tt SANC} library calculation can be modified by the user before it is 
loaded into our interface without intervention into the code of the interface itself.


\subsection {Conventions of Frames: KORALB, SANC and TAUOLA Interface}
\label{subsection:KORALBSANC}
\begin{figure}[h!]
\centering
\subfigure[KORALB-like orientation]{
\includegraphics[scale=0.25]{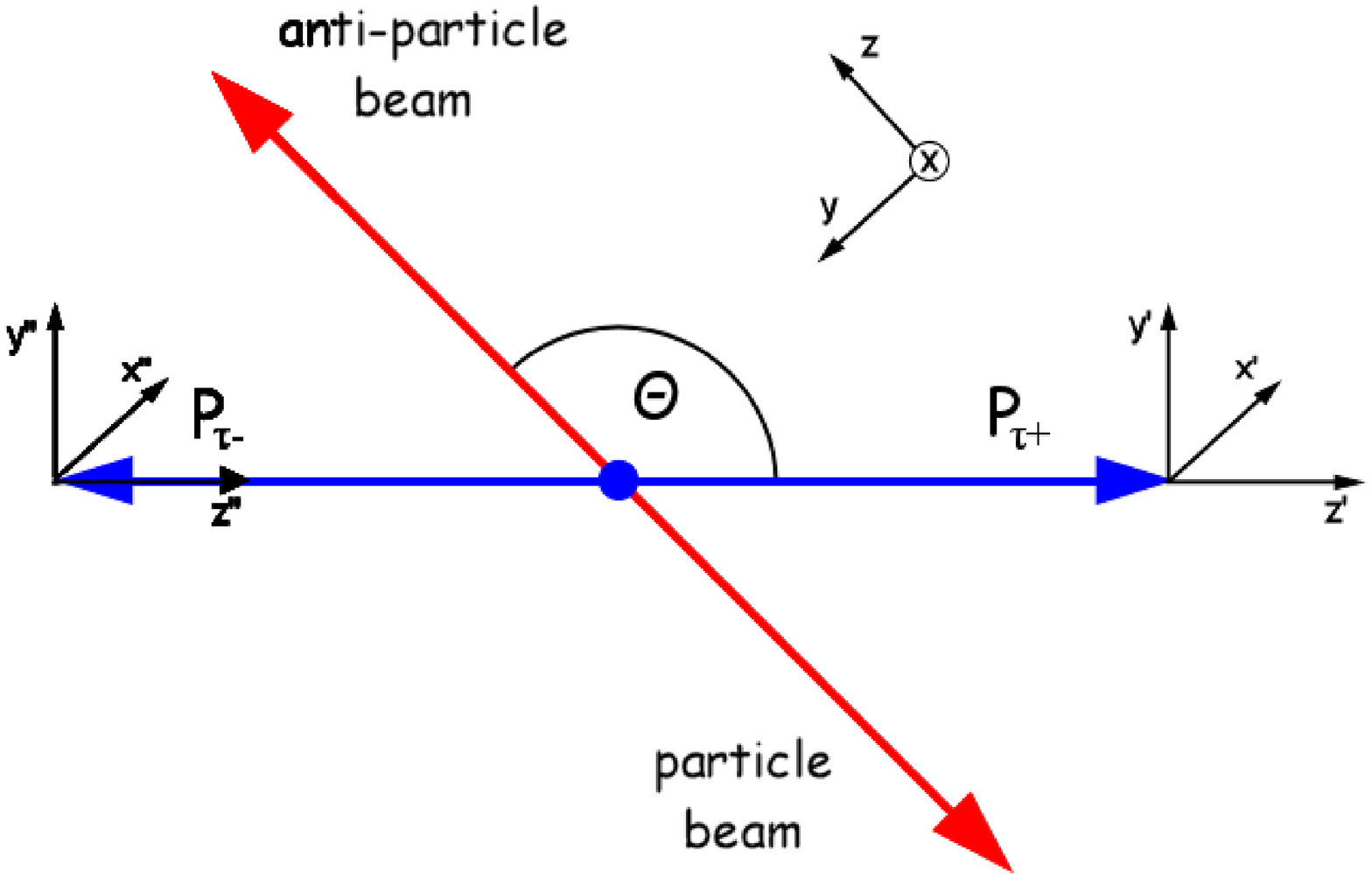}
\label{fig:designKB}
}
\subfigure[SANC module orientation]{
\includegraphics[scale=0.25]{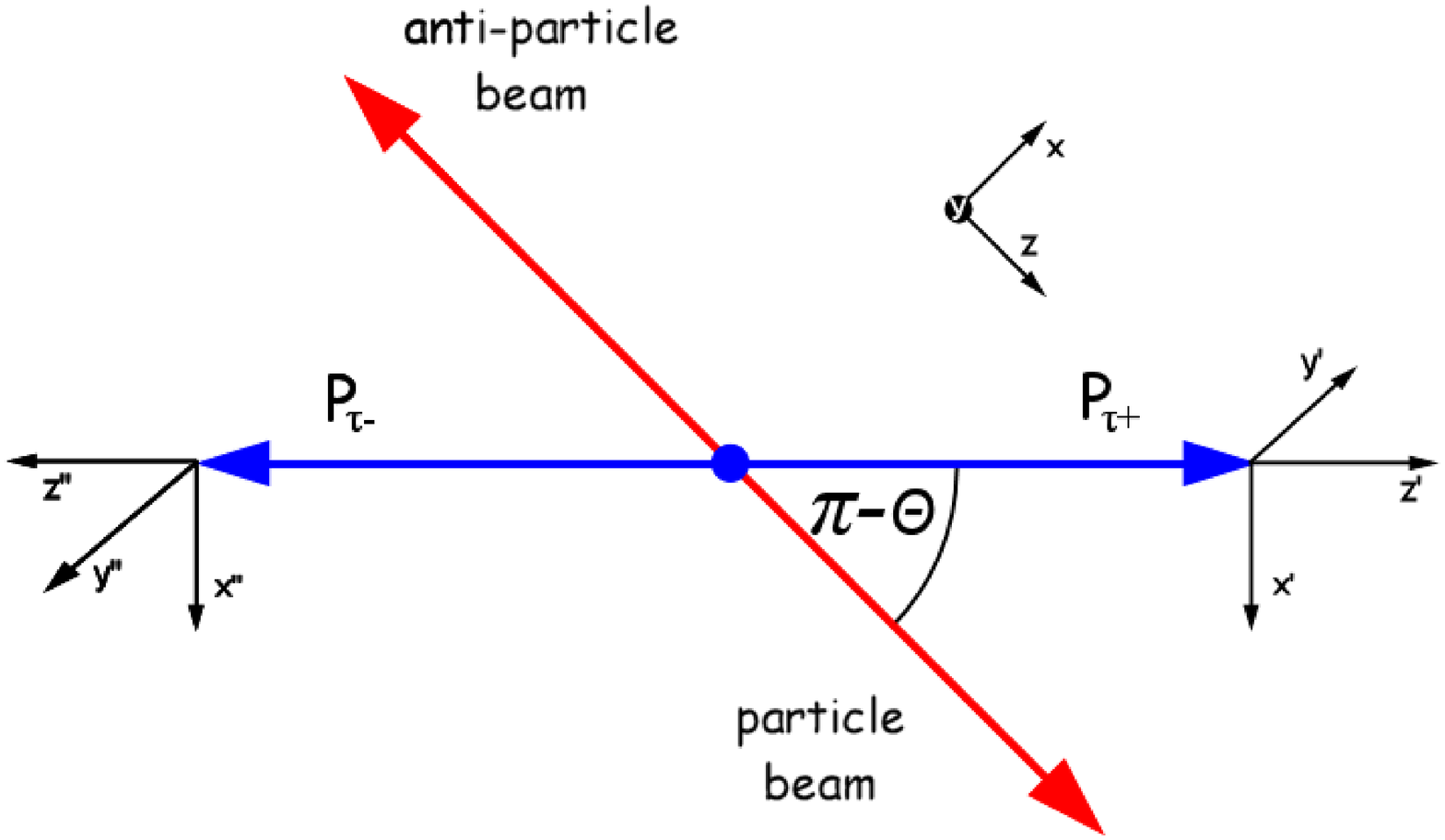}
\label{fig:designS}
}
\caption{
The relative orientation of reference frames for the spin states
of $\tau^+$, $\tau^-$ and for hard processes as used in our interface (Fig. (a)) and in {\tt SANC} module (Fig. (b)) are shown. In Fig (a) the axes $x$ (not shown explicitly), $x'$ and $x''$ 
are parallel to each other 
and  point behind the picture, axis $z$ is parallel to the direction  
of the anti-particle beam. In Fig (b) the axis $z$ is parallel to the direction of the particle beam, the axis 
$y'$ points behind the picture. Axes $y$ (not shown explicitly) and $y''$ point toward the reader and
are antiparallel to $y'$.  
}
\end{figure}

It is not essential for {\tt TAUOLA Interface}  and the  segment calculating electroweak 
corrections to follow exactly the same conventions for spin quantization. 
In {\tt TAUOLA Interface} we 
follow the frame orientation exactly as in  paper 
\cite{jadach-was:1984}. The adopted frame orientation is shown in Fig.~\ref{fig:designKB}.
In {\tt SANC}, the orientation of axes is different, see Fig.~\ref{fig:designS}
In the case of our interface, the beam momenta are laid along the z axis.  The anti-particle beam is parallel, particle beam is antiparallel.
The $y$ component of the $\tau^-$ is always positive.
The $\theta$ angle to be used for calculation of the density matrix is between
the directions of the antiparticle beam and the $\tau^+$.
In the case of the {\tt SANC} module, the $\tau$ momenta $p_{\tau^+}, p_{\tau^-}$ are laid on the $xz$ plane. The $xz$ plane is the reaction plane: a beam of particles (quarks or leptons) is parallel to the $z$ axis.
The $x$ component of $p_{\tau^-}$ is always negative.
The $y^{'}$ and $y^{''}$ axes of the $\tau^+$ and $\tau^-$ spin frames correspondingly have opposite direction to each other. The $y^{''}$ axis
is parallel to the $y$ axis of the hard process frame.
Appropriate rotations and other convention adjustments are performed by the program in preparation of the $R_{ij}$ tables: 
{\tt SANCtable.cxx}.

\subsection {Numerical Significance of Electroweak Corrections }
\begin{figure}[h!]
\centering
\includegraphics[scale=0.76]{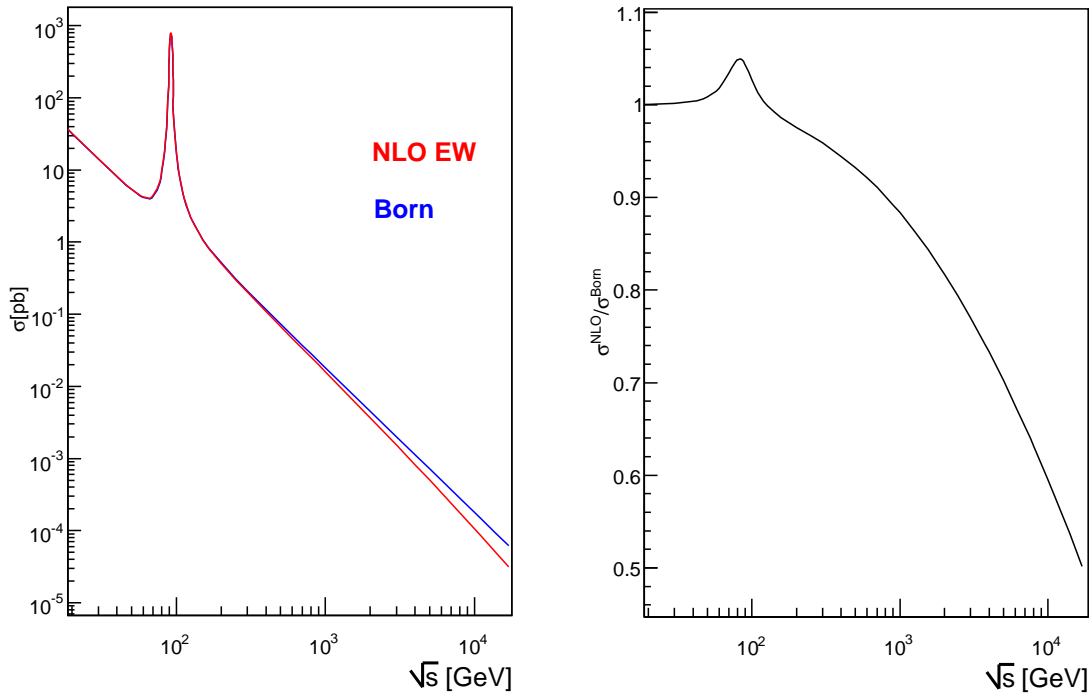}
\caption{\label{fig:sigmaUp} 
 The integrated cross section of $\tau$ pair production from up quarks  
calculated with and without NLO
EW corrections (red and blue lines) is shown in the left hand side plot. 
The ratio of the two distributions 
is given on the right hand plot.   
We are using the alpha scheme for electroweak corrections. That is why 
light fermion loops contribute to the difference between the two lines.
 The differences between the alpha scheme Born predictions  and expressions
used in the host program
must be understood before the correcting weight (see Appendix \ref{subsection:eHelicity-ew}) is used.
 }
\end{figure}

\begin{figure}[h!]
\centering
\includegraphics[scale=0.76]{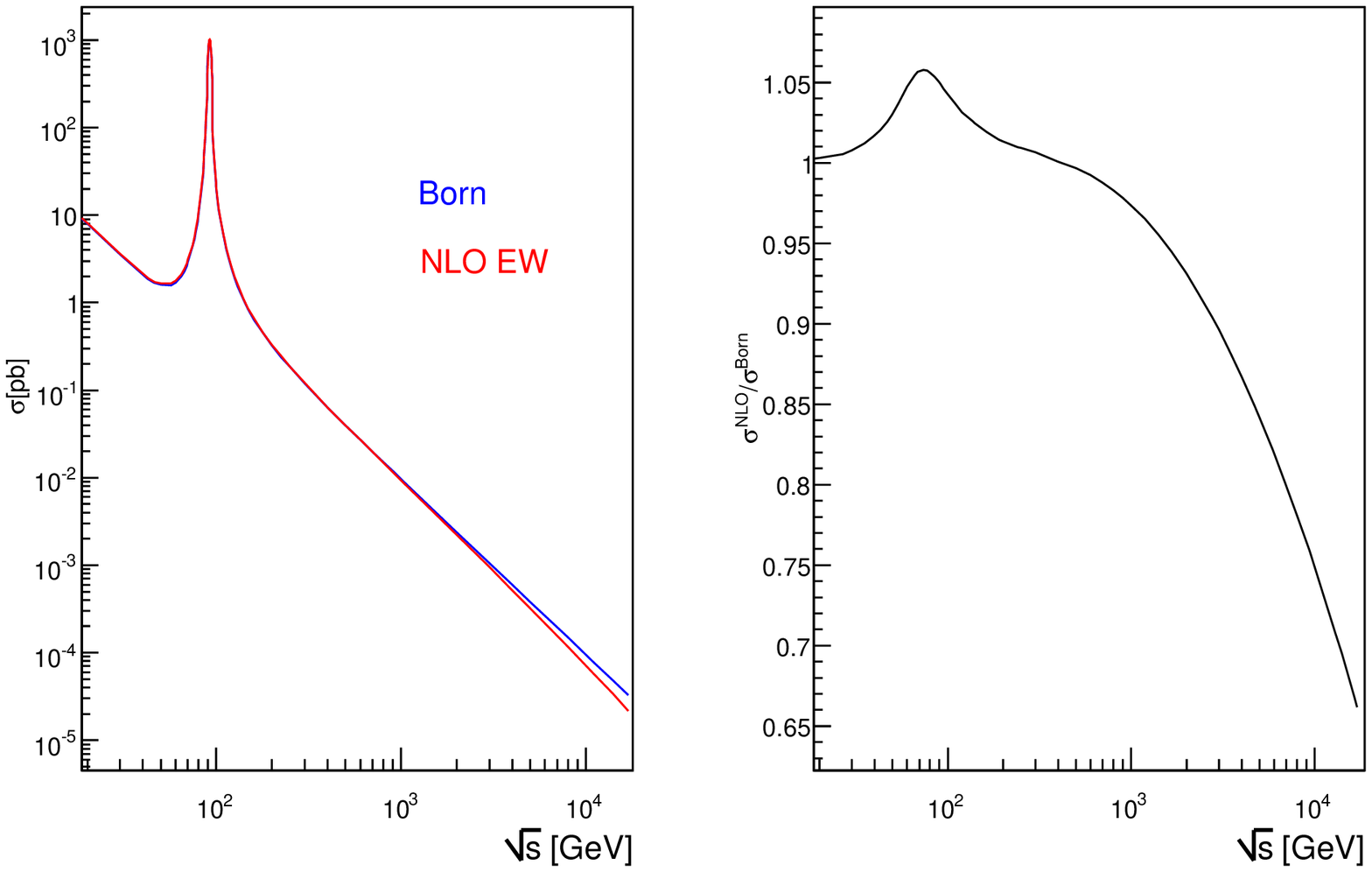}
\caption{\label{fig:sigmaDown} 
 The integrated cross section of $\tau$ pair production from down quarks  
calculated with and without NLO
EW corrections (red and blue lines) is shown in the left hand side plot. 
The ratio of the two distributions 
is given on the right hand plot.   
We are using the alpha scheme for electroweak corrections. That is why 
light fermion loops contribute to the difference between the two lines.
 The differences between the alpha scheme Born predictions  and expressions
used in  host program
must be understood before the correcting weight (see Appendix \ref{subsection:eHelicity-ew}) is used.
 }
\end{figure}

\begin{figure}[h!]
\centering
\subfigure[Up quarks]{
\includegraphics[scale=0.35]{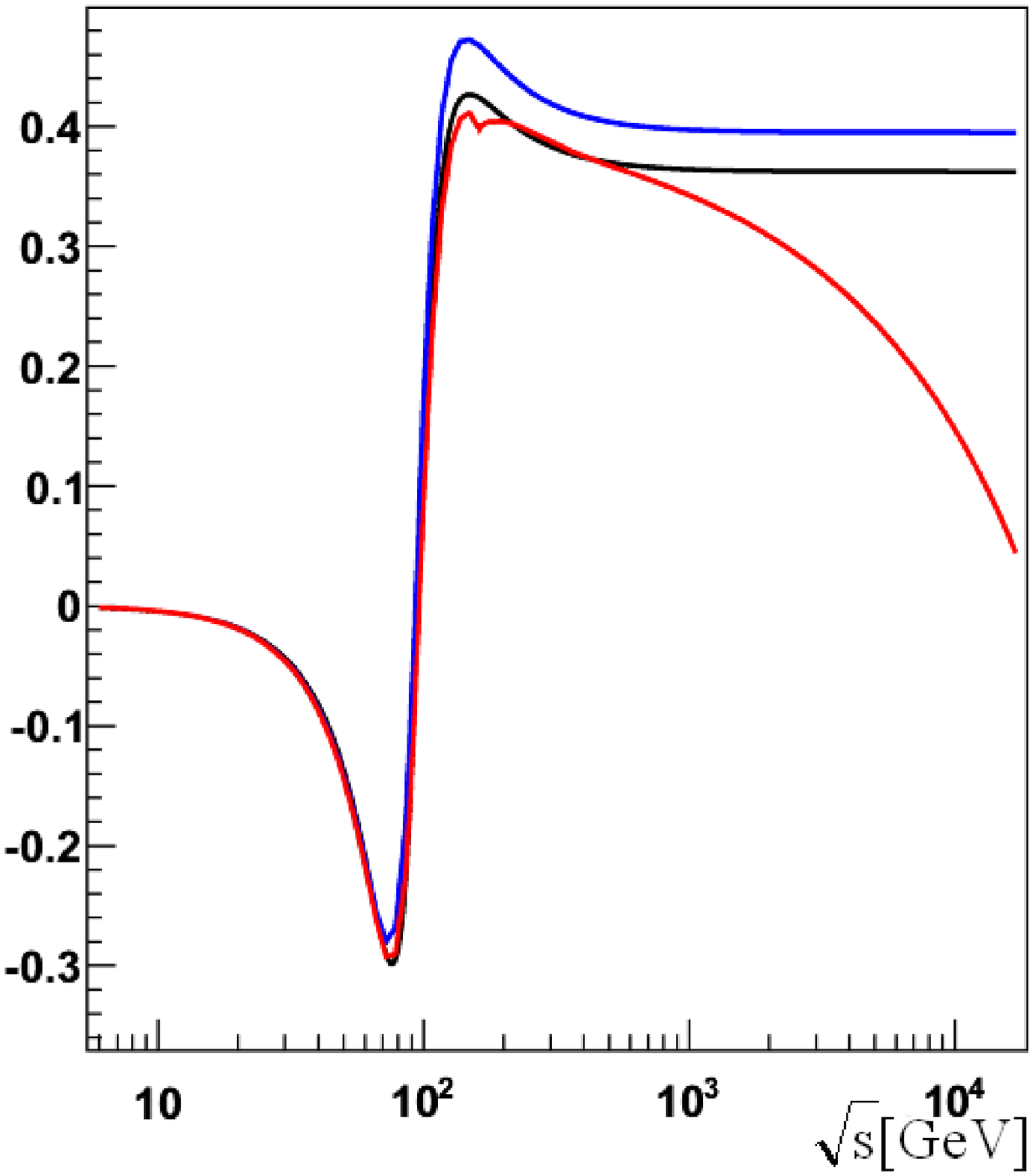}
\label{fig:polUp}
}
\subfigure[Down quarks]{
\includegraphics[scale=0.35]{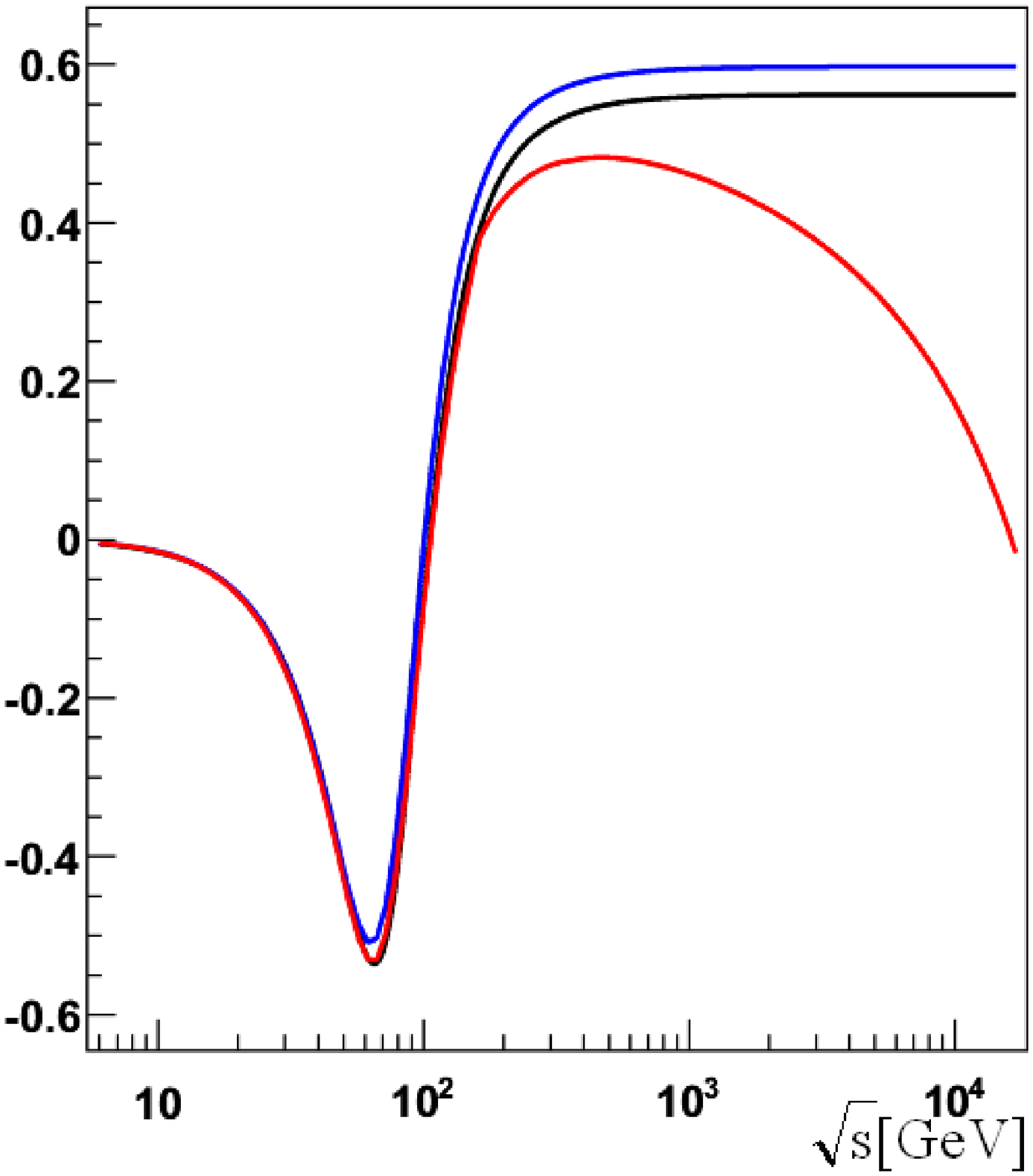}
\label{fig:polDown}
}
\caption{ Polarization for $\tau$ leptons produced from up quarks
  (Fig. (a)) and down quarks (Fig. (b)) at $\cos\theta=-0.2$.  The red
  line is with electroweak corrections, the black is Standard Born as
  is default in the interface. The blue line is Born according to
  alpha scheme.  The main purpose of these results is a technical test
  of the software installation.  Note however the inadequateness of the alpha
  scheme Born, which is significantly different from the other two
  results even at relatively low energies. The small bump on the red
  line on Fig. (a) is due to the WW threshold. It is insignificant for
  positive $\cos\theta$.  }
\end{figure}

One may wonder whether the numerical results induced by electroweak corrections are
of any practical purpose. They are expected to be of the order of
1\%  and indeed are not that large  for 
the intermediate state virtuality of up to 100 GeV above the $Z$ boson mass.
The situation changes however significantly at higher energies. As can be 
seen from  Figures \ref{fig:sigmaUp} and  \ref{fig:sigmaDown} the effect may 
be of the order of even 50\% at virtualities of several TeV. 
This is quite in agreement with the results of refs. 
\cite{Adam:2008ge,Adam:2008pc}. In  Figures 
\ref{fig:polUp} and  \ref{fig:polDown} we collect results for 
$\tau$ polarization calculated at $\cos \theta=-0.2$. Again,
 the effects are small up to the energy scale of about 500 GeV. At larger scales
 corrections  become sizable. The electroweak corrections
should be therefore considered in studies aiming for new physics phenomena
such as $Z'\to \tau^+\tau^-$ decays.

\section{Tests  of Spin Correlations and Numerical Results}

There are two purposes of the presented results in this section. On
one hand these results complement the technical tests described in
Appendix ~\ref{sect:elem} with the ones oriented toward a particular
hard processes.  The technical tests should be repeated for every new
program installation or configuration. On the other hand, results of
the present section are of potential physics interest as well. They
illustrate the dominant spin effects on idealized distributions for LHC
measurements.

\label{sec:results}

Tests presented here
were conducted using {\tt MC-TESTER}\cite{Golonka:2002rz,Davidson:2008ma}. 
{\tt MC-TESTER} allows semi-automated comparisons of invariant mass 
distributions of each sub-group of eg. $\tau$ or $Z$ stable decay products.  
The results of these tests were also compared to the 
 results obtained  with  the {\tt FORTRAN  Interface} (which 
has been well validated by comparison with analytical and numerical
calculations for $\tau$ pair production\footnote{This represents
tests of interface. In all cases $\tau$ decays are generated 
with the help of {\tt TAUOLA FORTRAN}.
For a review of physics oriented tests of $\tau$ decays themselves, and 
projects for future improvements based on low energy $e^+e^-$ data, 
see ref. \cite{Actis:2009gg}.}). 

In addition to this, we created custom {\tt MC-TESTER } 
macros for plotting other spin sensitive quantities and compared these to 
published results. Numerical results are presented later in the section, see Figs.~\ref{fig:H_spin_3}, \ref{fig:H_spin_4} and \ref{fig:H_spin_C}, \ref{fig:H_spin_D}.


\subsection{$Z/\gamma \rightarrow \tau^+ \tau^-$}
\label{mctester_results_z}

\begin{figure}[h!]
\centering
\subfigure[$M_{\pi^+\pi^-}$]{
\includegraphics[scale=0.35]{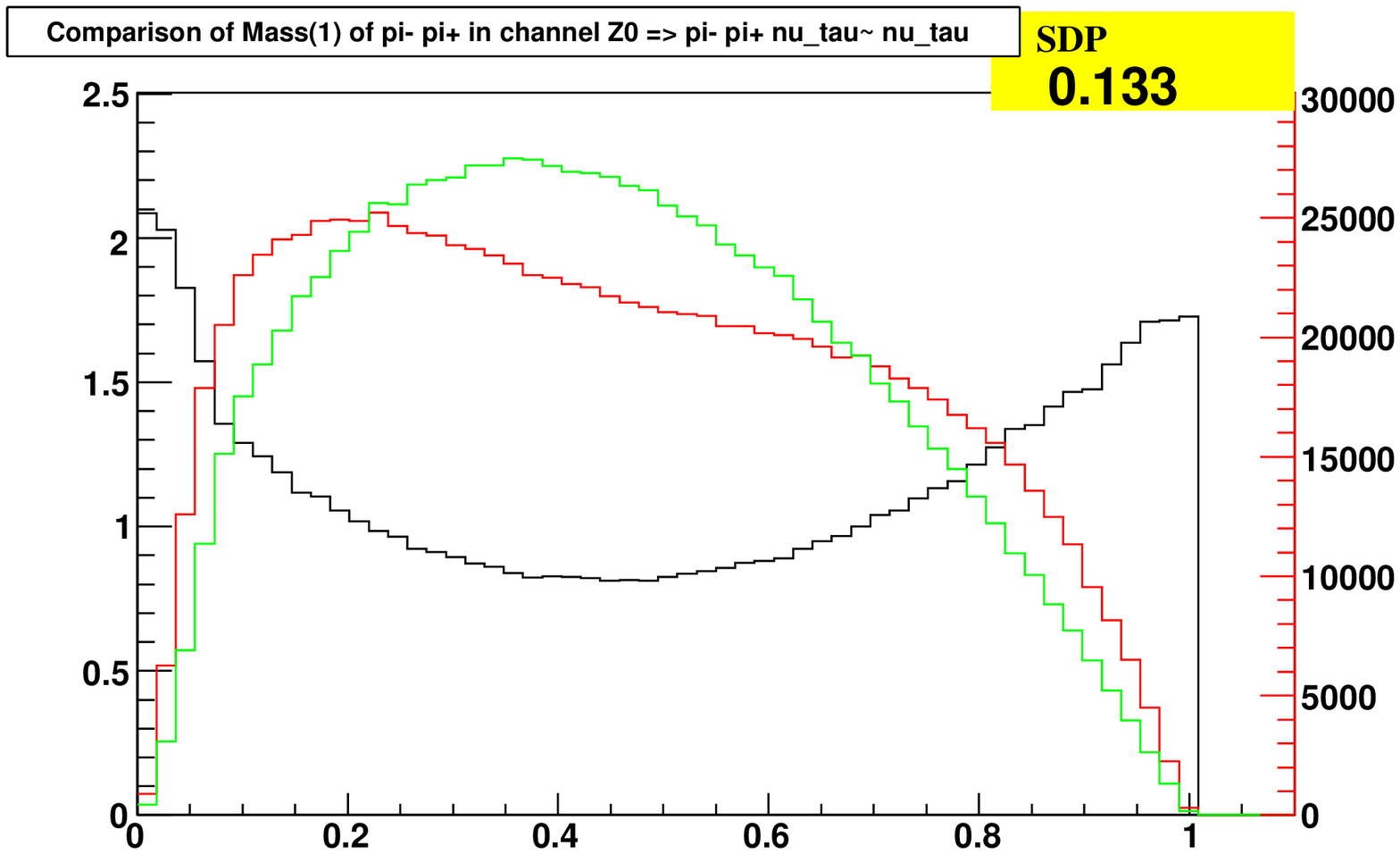}
\label{fig:Z_spin_1}
}
\subfigure[$1-2\frac{E_{\pi^+}}{M_Z}$]{
\includegraphics[scale=0.35]{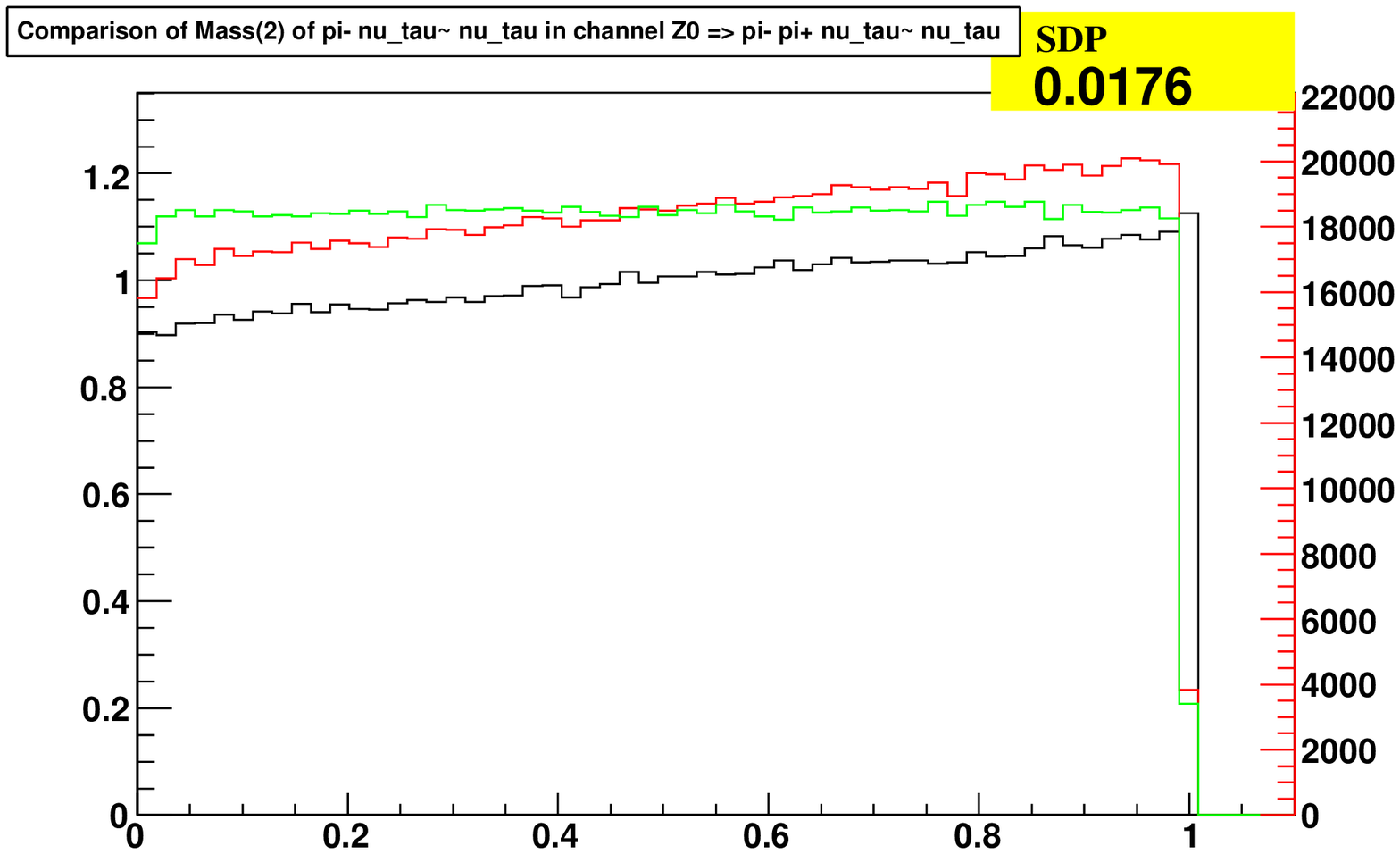}
\label{fig:Z_spin_2}
}
\caption{Longitudinal spin observables for the Z boson ($e^+e^- \rightarrow Z$ at 500 GeV).
Distributions are shown for spin effects switched on (red), spin effects switched off (green), and their ratio (black).
}
\end{figure}

The longitudinal spin effects for $Z$ decay into $\tau$'s was tested
by restricting the $\tau$ decay mode to $\tau^{\pm} \rightarrow \pi^{\pm} \nu_{\tau}$
and examining the invariant mass of the $\pi^+\pi^-$ pair, $M_{\pi^+\pi^-}$ 
(see Fig. \ref{fig:Z_spin_1}) and the $\pi$ energy distribution in the rest frame of 
the $Z$ (see Fig. \ref{fig:Z_spin_2}). The effect of $Z$ polarization on these distributions 
was studied in \cite{Pierzchala:2001gc} and we obtained consistent results with the new 
C++ implementation of the {\tt TAUOLA Interface}.

\subsection{$H^0/A^0 \rightarrow \tau^+ \tau^-$}
\label{subsection:HA}
\begin{figure}[h!]
\centering
\subfigure[$M_{\pi^+\pi^-}$]{
\includegraphics[scale=0.35]{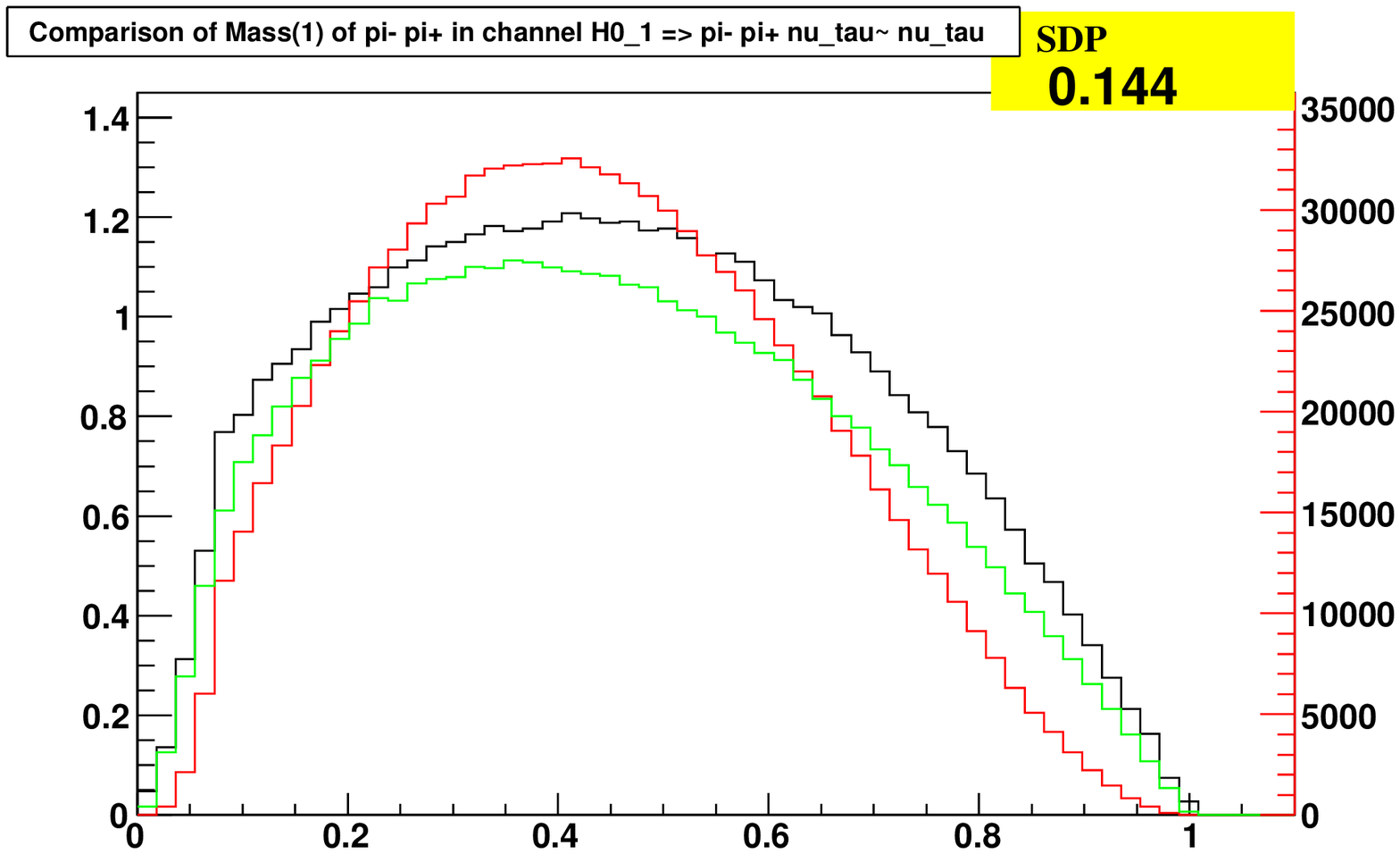}
\label{fig:H_spin_1}
}
\subfigure[$1-2\frac{E_{\pi^+}}{M_Z}$]{
\includegraphics[scale=0.35]{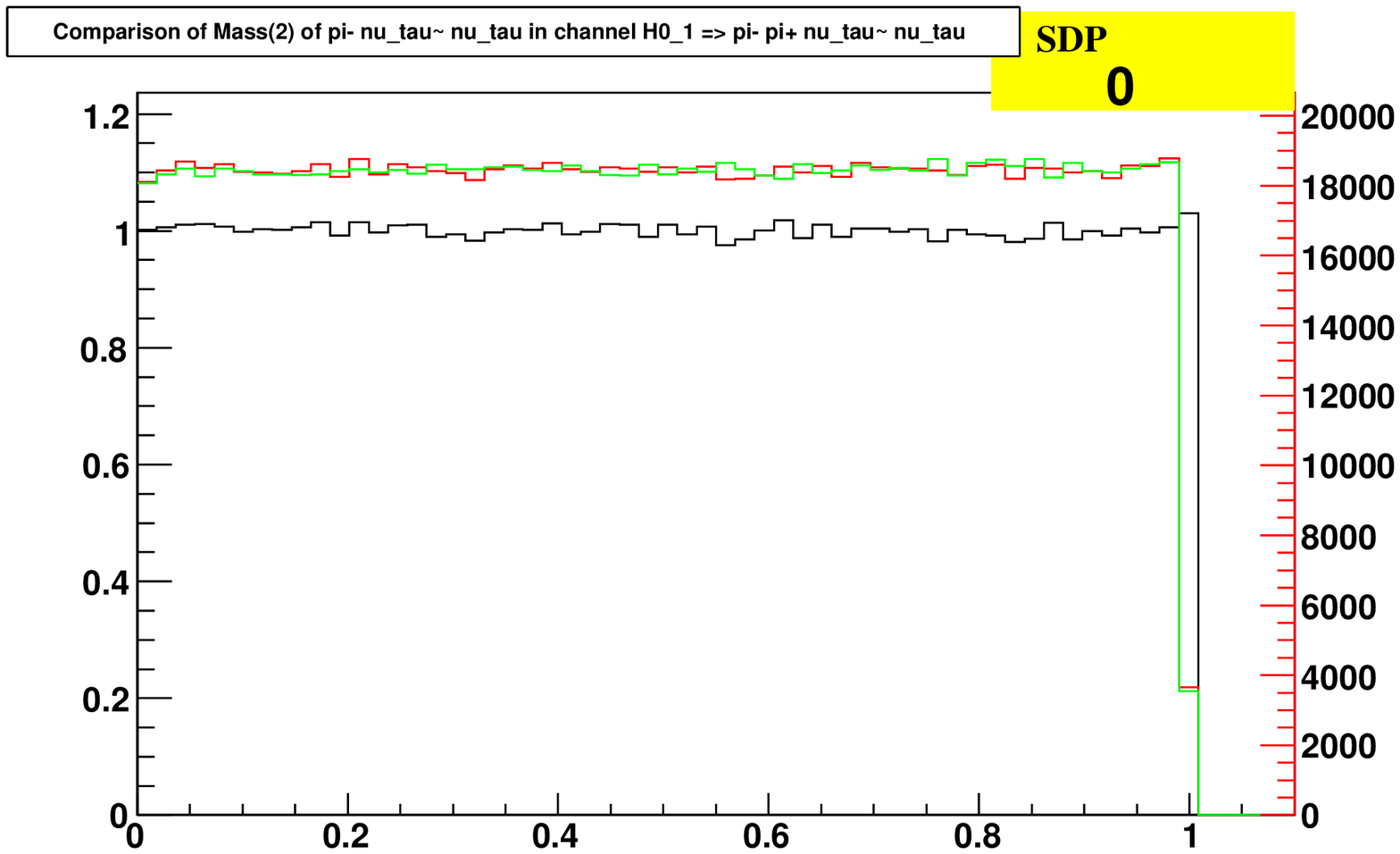}
\label{fig:H_spin_2}
}
\caption{Longitudinal spin observables for the H boson for $\tau^{\pm}\rightarrow \pi^{\pm} \nu_{\tau}$.
Distributions are shown for spin effects switched on (red), spin effects switched off (green), and their ratio (black).  }
\end{figure}

As was done for $Z$ decay in Section \ref{mctester_results_z}, longitudinal spin 
effects for Higgs decay into $\tau$'s was tested using $M_{\pi^+\pi^-}$ (Fig. \ref{fig:H_spin_1}) 
and the $\pi$ energy distribution in the rest frame of the $H^0$ (see Fig. \ref{fig:H_spin_2}),
which was flat as expected.

\begin{figure}[h!]
\centering
\subfigure[$\pi^+ \pi^-$ acollinearity distribution ($\approx \pi$)) ]{
\includegraphics[scale=0.35]{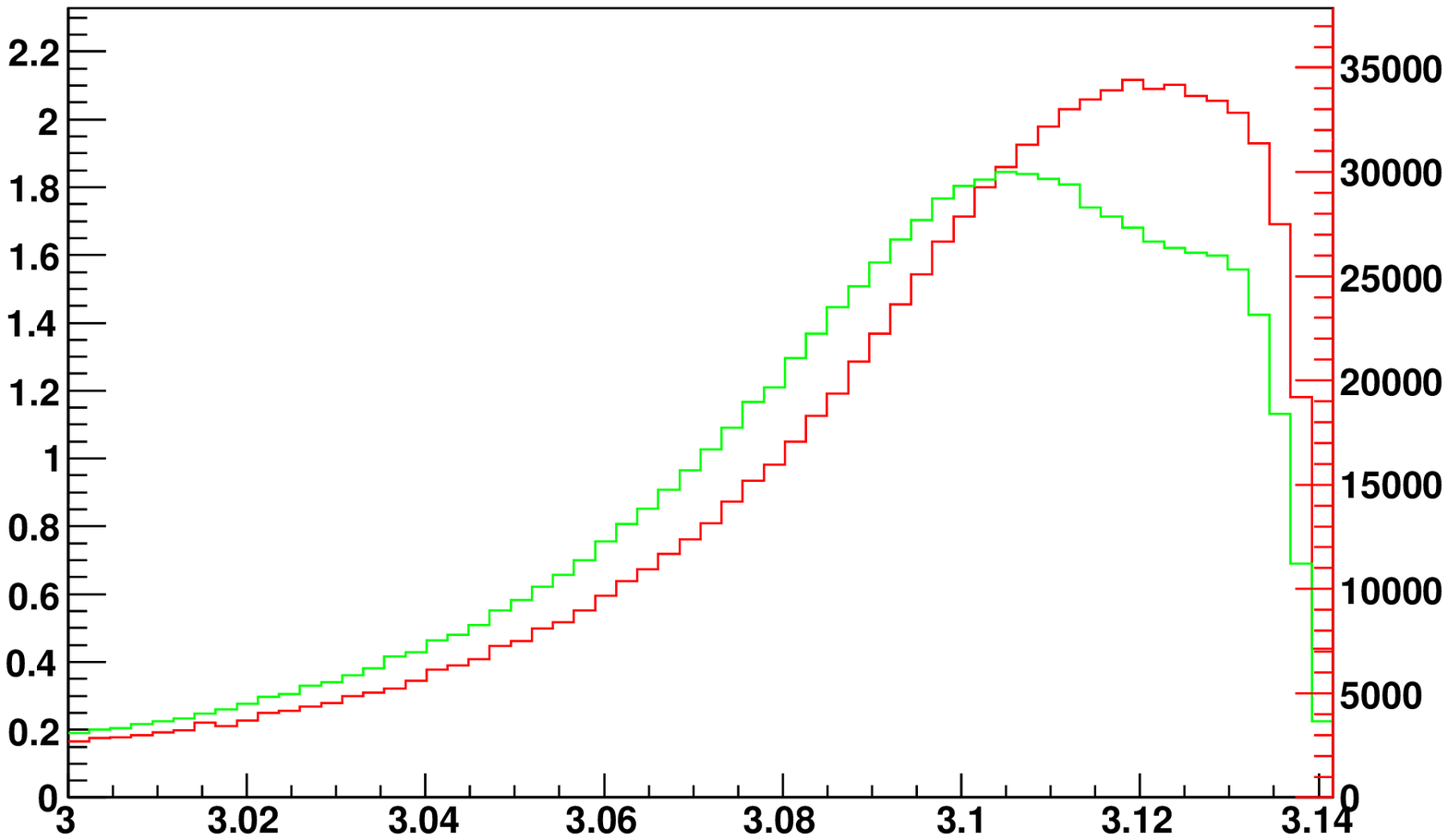}
\label{fig:H_spin_3}
}
\subfigure[$\pi^+ \pi^-$ acoplanarity distribution]{
\includegraphics[scale=0.35]{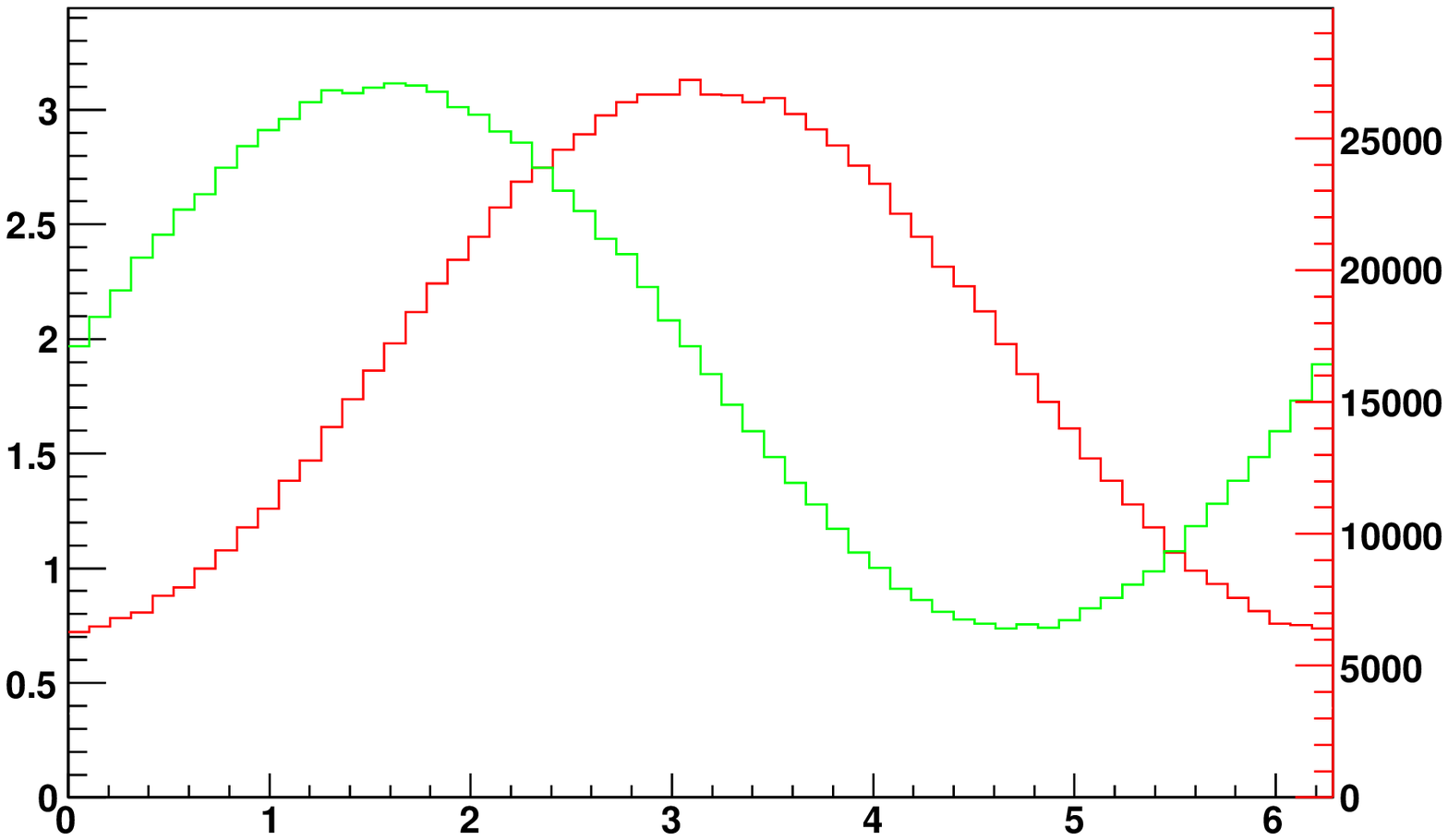}
\label{fig:H_spin_4}
}
\caption{Transverse spin observables for the Higgs boson for $\tau^{\pm}\rightarrow \pi^{\pm} \nu_{\tau}$.
Distributions are shown for scalar Higgs (red), scalar-pseudoscalar Higgs with mixing angle $\frac{\pi}{4}$. For the definition of angles see Section \ref{subsection:HA}.}
\end{figure}

Let us now turn  to transverse spin correlations. In Fig. \ref{fig:H_spin_3}
the benchmark histogram as produced by our {\tt FORTRAN Interface} and given in Fig.3 of 
reference \cite{Was:2002gv} is reproduced\footnote{  In the plot
the case of non zero scalar-pseudoscalar mixing was chosen. This is the origin of the difference with
ref. \cite{Was:2002gv}.}. 
It features acollinearity of the $\pi^+$, $\pi^-$ pair in
the Higgs boson rest frame, both $\tau$'s  decay to $\pi^\pm \nu$.
For the same decay set up, Fig. \ref{fig:H_spin_4} features acoplanarity of the planes built respectively on decay products of $\tau^+$ and $\tau^-$. The spin effect is indeed large. However, it requires
use of unobservable neutrino momenta. It is difficult or even impossible 
to achieve sufficient experimental precision in reconstruction of the reaction frame necessary for this
observable. Also, the first observable presented on 
Fig. \ref{fig:H_spin_3} suffers from the same limitation.

\begin{figure}[h!]
\centering
\subfigure[acoplanarity distribution ($y_1y_2>0$)]{
\includegraphics[scale=0.35]{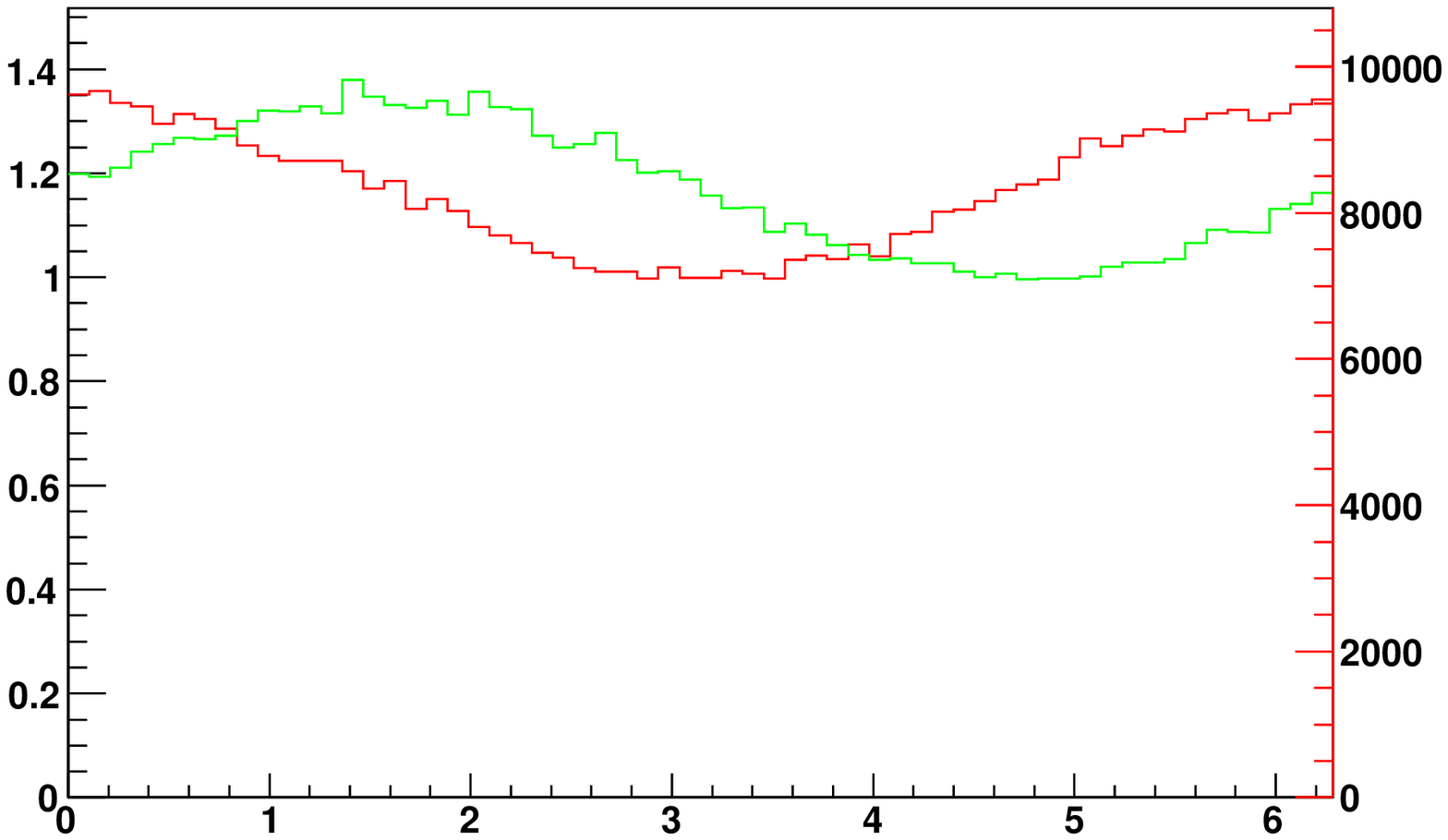}
\label{fig:H_spin_C}
} 
\subfigure[acoplanarity distribution ($y_1y_2<0$)]{
\includegraphics[scale=0.35]{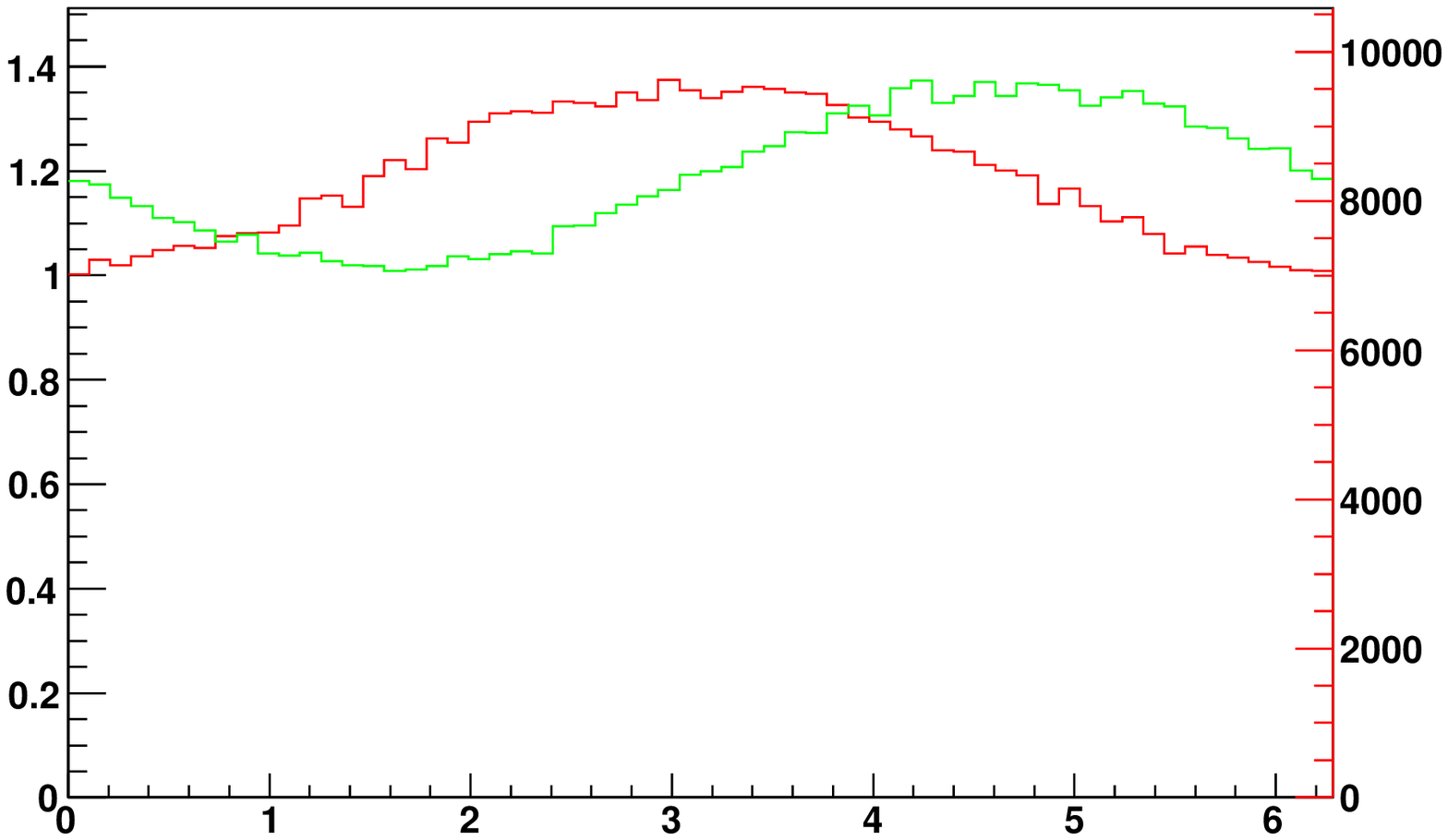}
\label{fig:H_spin_D}
}
\caption{Transverse spin observables for the Higgs boson for 
$\tau^{\pm}\rightarrow \pi^{\pm} \pi^0 \nu_{\tau}$.
Distributions are shown for scalar Higgs (red), scalar-pseudoscalar 
Higgs with mixing angle $\frac{\pi}{4}$ (green). 
For the definition of angles see Section \ref{subsection:HA}.}
\end{figure}

The two other tests, Figures \ref{fig:H_spin_C}  and \ref{fig:H_spin_D} present distribution of acoplanarity
angle for  the two planes built respectively on the momenta of $\pi^+\pi^0$ and 
$\pi^-\pi^0$; the decay products of $\rho^+$ and 
$\rho^-$. All in the rest frame of the $\rho$-pair. It is directly based
on measurable quantities. The $\rho^\pm$  originate respectively 
from $\tau^\pm \to \nu \rho^\pm$ decays.
There is no need for Higgs rest frame reconstruction in this case. Events are divided
into two categories. If the energy difference between
charged and neutral pions coming from the two $\tau$'s are of the same sign, they
contribute to Fig. \ref{fig:H_spin_C},
otherwise they contribute to  Fig. \ref{fig:H_spin_D}. For details of the 
definition and for more numerical results obtained with the {\tt FORTRAN} Interface, 
see \cite{Desch:2003rw}.

\subsection{$W^{\pm} \rightarrow \tau^{\pm} \nu_{\tau}$ and $H^{\pm} \rightarrow \tau^{\pm} \nu_{\tau}$}
\label{sect:WH}

\begin{figure}[h!]
\centering 
\subfigure[$W^+ \to \nu_\tau \tau^+, \; \tau^+ \to \pi^+ \bar \nu_\tau$]{
\includegraphics[scale=0.35]{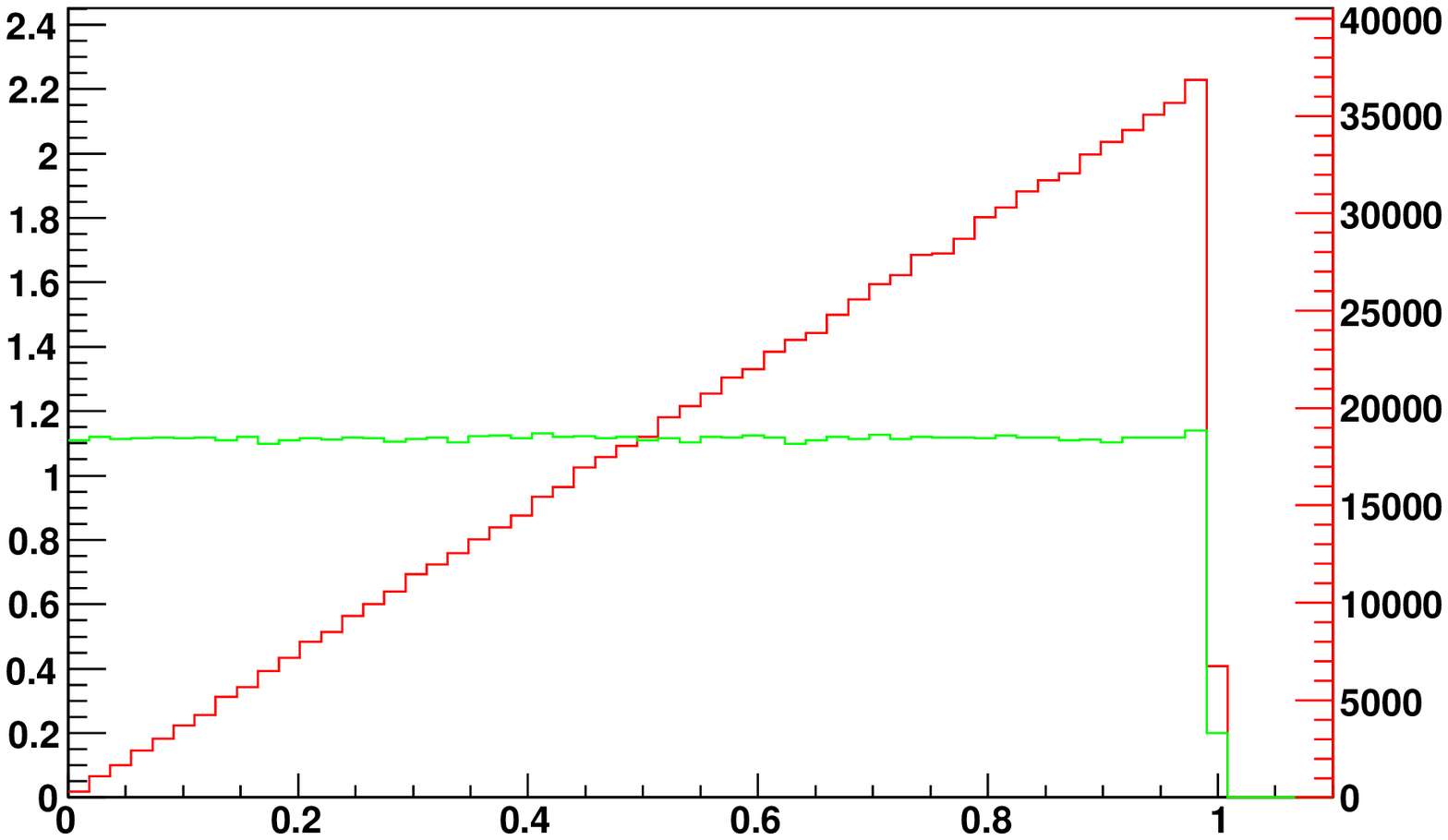}
\label{fig:W}
}
\subfigure[$H^+ \to \nu_\tau \tau^+, \; \tau^+ \to \pi^+ \bar \nu_\tau$]{
\includegraphics[scale=0.35]{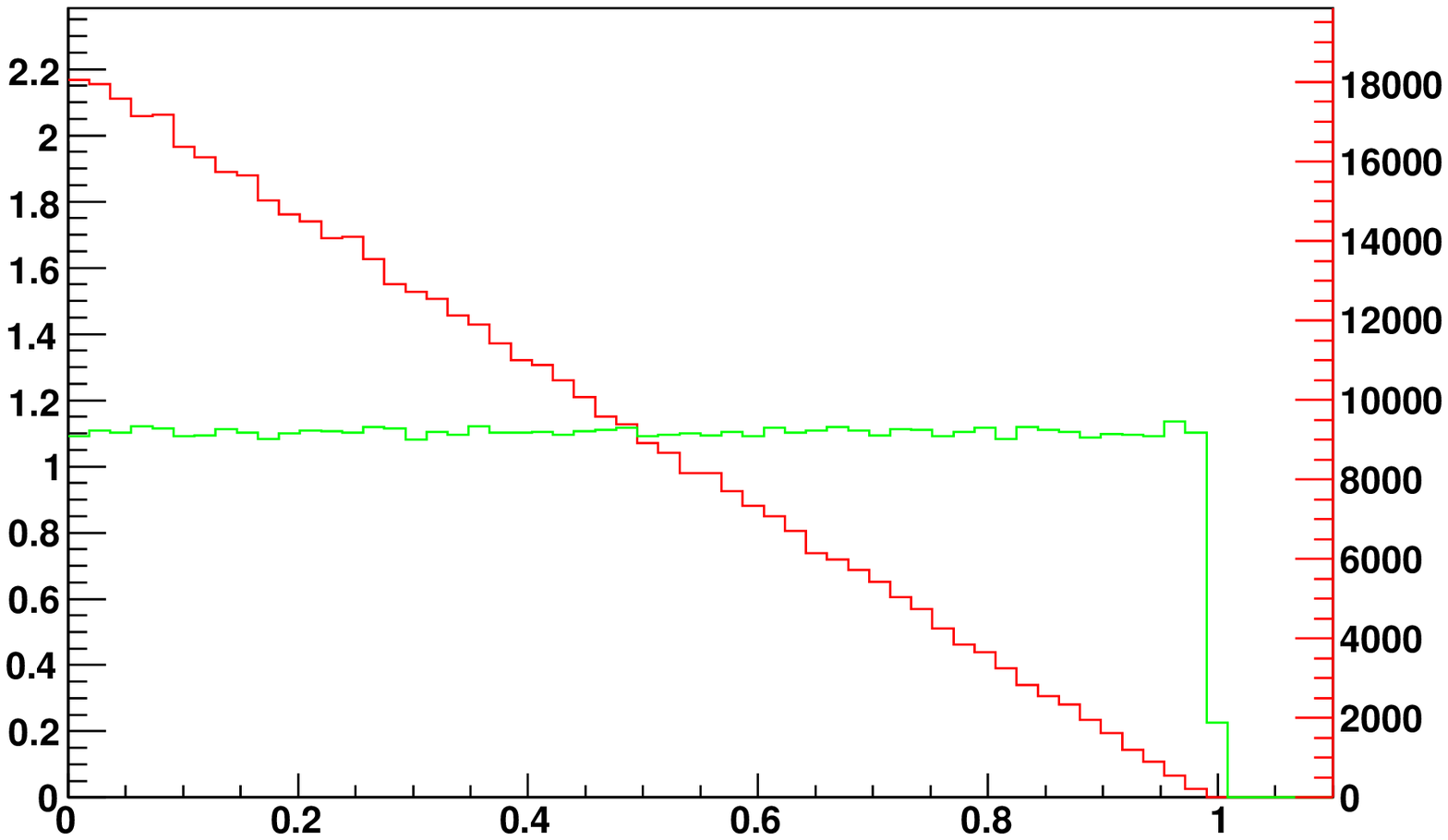}
\label{fig:H}
}\caption{ 
 Pion energy
spectrum in the rest frame of $W$ (left hand side) and $H^+$ (right-hand side).  
Spin effects included (red line) and neglected
(green line) are plotted.   The variable $1-2\frac{E_{\pi^+}}{M_{W^+}}$ or
$1-2\frac{E_{\pi^+}}{M_{H^+}}$ is  used respectively.}
\end{figure}

For the simplest decay mode $\tau^{\pm}\rightarrow \pi^{\pm} \nu_{\tau}$,  
as was already discussed in ref.~\cite{Pierzchala:2001gc}, 
the pion energy spectrum
should be softer  in the case of $W^\pm$ decays and harder in the case of charged Higgs decay.
This is indeed reproduced in Figs. \ref{fig:W} and \ref{fig:H}
and  the spectra are reversed for the two cases.

\newpage

\section{Outlook}
Let us summarise briefly the next steps which are planned for the work on TAUOLA
Universal Interface.
Further extension of our work will be focused on generating spin states 
in the production processes from quantities which can be 
experimentally measured for real data events. Discussion of 
systematic error for such reconstructed spin states will require discussion of QCD corrections. 

At present, our interface is designed to work with the {\tt HepMC} event record. 
However, following the design we tested in 
{\tt MC-TESTER} \cite{Golonka:2002rz,Davidson:2008ma} it will be rather easy 
to adapt to any other event structure, by writing the appropriate event record 
interface.

{\tt TAUOLA Universal Interface} as well as {\tt TAUOLA} itself are expected to remain 
framework-like code where the user is supposed to modify some of the parts according
to her/his particular purposes. 

The segment of code for analyzing the hard process and generating spin
states is now becoming a significant component of the project and
already exceeds by far the category of peripheric methods related to
the {\tt TAUOLA Interface}. In the future it should be moved to a
separate class.

We expect that in the next few years  better parametrization of 
hadronic form-factors based on $\tau$ data from the Belle and BaBar collaborations
and refined models of decays will become available. Then the directory 
{\tt /tauola-fortran} will be 
replaced by the new version incorporating those achievements.

\vskip 5 mm
\centerline{\Large\bf Acknowledgments}
\vskip 5 mm

Useful discussions with P. Golonka during the early stage of project development and discussions 
with members of the ATLAS and CMS collaborations, and the LCG group are acknowledged.
We are specially indebted to the pilot users of the interface, in particular to: Julia Yarba, Sami Lehti, Eric Torrence, 
Marcin Wolter
and Anna Kaczmarska.

Partial support of Polish-French collaboration
no. 06-124 within IN2P3 through LAPP Annecy during final completion of this work is
also acknowledged.

\providecommand{\href}[2]{#2}

\newpage
\appendix

\section{Appendix: Interface to {\tt TAUOLA FORTRAN}}
\label{Interface to TAUOLA}

From the point of view of a C++ interface, the $\tau$ decay library should be 
seen as a black-box. The code of the library is stored in the  directory  {\tt tauola-fortran} and 
is 
identical to the one stored in the directory {\tt TAUOLA}, documented in ref. \cite{Golonka:2003xt}. 
Minor adaptations which have been made affect 
 platform-dependent files only. In the future, we expect this part 
of the code to be replaced
with the new version based on work proposed in \cite{Actis:2009gg}. 
This is one of the main reasons why we still keep the 
code  in FORTRAN.

This section is addressed to developers of the interface, 
and special users interested in versions of the {\tt TAUOLA} initialization
other than default one, {\tt cleo}. For this purpose
we describe the common blocks and routines which allow
communication between {\tt TAUOLA} and {\tt TAUOLA C++  Interface}.  Even though
they are explained already in
refs~\cite{Jadach:1990mz,Jezabek:1991qp,Jadach:1993hs,Golonka:2003xt}.
The repetition here is convenient for easier explanation of user configuration discussed in Appendix
\ref{sec:User Configuration}.

\subsection{Common Blocks}

\begin{description}
\item[IDFC] $\tau$ PDG id
    \begin{description}
    \item[IDFF] \textit{int}  PDG id of the `first' $\tau$ must be 15 or -15
    \end{description}
\end{description}

\begin{description}
\item[TAUPOS] Position of $\tau$'s in the event record common block
    \begin{description}
    \item[NPA] \textit{int} first $\tau$ position
    \item[NPB] \textit{int} second $\tau$ position
    \end{description}
\end{description}

\begin{description}
\item[PARMAS] Particles masses and widths
  \begin{description}
  \item[AMTAU] \textit{float}  mass of $\tau$
  \item[AMNUTA] \textit{float}  mass of $\nu_\tau$
  \item[AMEL] \textit{float}  mass of $e$
  \item[AMNUEL] \textit{float} mass of $\nu_e$
  \item[AMMU] \textit{float} mass of $\mu$
  \item[AMNUMU] \textit{float} mass of $\nu_\mu$
  \item[AMPIZ] \textit{float} mass of $\pi^0$
  \item[AMPI] \textit{float} mass of $\pi^\pm$ 
  \item[AMRO] \textit{float}  mass of $\rho$ (As used in some but not all
decay channels for parametrization of hadronic currents. The same is true for all other unstable intermediate state particles and resonances.)
  \item[GAMRO] \textit{float} width of $\rho$
  \item[AMA1] \textit{float} mass of $a_1$ 
  \item[GAMA1] \textit{float} width of $a_1$
  \item[AMK] \textit{float} mass of $K^\pm$
  \item[AMKZ] \textit{float} mass of $K^0$
  \item[AMKST] \textit{float} mass of $K^*$
  \item[GAMKST] \textit{float} width of $K^*$
    \end{description}
\end{description}    

\begin{description}
\item[JAKI] control variables for the decay channels
  \begin{description}
  \item[JAK1] \textit{int} chosen decay channel for the first $\tau$, if set to $0$
a random choice will be done accordingly to predefined branching ratios
  \item[JAK2] \textit{int}
chosen decay channel for the second $\tau$, if set to $0$
a random choice will be done accordingly to predefined branching ratios
  \item[JAKP] \textit{int} used in some FORTRAN applications only
  \item[JAKM] \textit{int} used in some FORTRAN applications only
  \item[KTOM] \textit{int} used in some FORTRAN applications only
  \end{description}
\end{description}  

\begin{description}
\item[TAURAD] variables for QED radiative corrections in leptonic $\tau$ decay channels
    \begin{description}
      \item[ITDKRC] \textit{float} 1/0 radiative correction on/off
      \item[XK0DEC] \textit{float} minimal energy of photon to be generated
(with respect to maximum possible value).
  \end{description}
\end{description}  

\begin{description}
\item[TAUBRA] Variables for composition of $\tau$ decay channels. 
    \begin{description}
      \item[{GAMPRT[30]} ] \textit{float (non negative)} one dimensional matrix of $\tau$ decay branching ratios used if {\tt JAK1=0} or {\tt JAK2=0}. Can be changed by the user at any time of generation. GAMPRT does not need to sum up to 1. Default of any entry can be changed by invoking 
method  {\tt Tauola::setTauBr(int i, double br)}. 

      \item[{JLIST[30]} ] \textit{integer} one  dimensional table, to be left unchanged. It is basically a FORTRAN emulation of a table of pointers.
      \item[NCHAN] \textit{integer} number of $\tau$ decay channels which can be used, all higher than NCHAN values of GAMPRT are dummy.  
  \end{description}
\end{description}

\begin{description}
\item[TAUKLE]  Further variables for composition of some of the $\tau$ decay sub-channels
  \begin{description}
  \item[BRA1] \textit{float} relative branching ratio between  $a_0 \to \pi^+\pi^+\pi^-$ and $a_0 \to \pi^0\pi^0\pi^+$
  \item[BRK0] \textit{float} relative branching ratio of $ K_0$ decay
  \item[BRK0B] \textit{float} relative branching ratio of $\bar K_0$ decay
  \item[BRKS] \textit{float} relative branching ratio of $ K^*$ decay
  \end{description}
\end{description}  

\subsection{Routines}
\label{subsubsec:routines}
  
\begin{description}
\item[INIETC] Initialize content of JAKI and TAURAD common blocks. \\
  Return type: \textit{void} \\
  Parameters: \textit{none}

\end{description}

\begin{description}
\item[INIMAS] Initializes masses stored in common block PARMAS \\
  Return type: \textit{void} \\
  Parameters: \textit{none}
\end{description}

\begin{description}
\item[INIPHX] Initializes parameters of QED, common block QEDPRM. \\
  Return type: \textit{void} \\
  Parameters: \textit{none}

\end{description}

\begin{description}
\item[INITDK] Initializes kinematical information on $\tau$ decay channels: 
branching fractions to be used for JAK1,JAK2=0, masses and flavours of scalars
for each decay channel, names of the decay channels. It is well documented
in the FORTRAN version of the program. This routine should be left unchanged.
Defaults of the GAMPRT matrix residing in the common block TAUBRA, can be changed
before any consecutive execution of the routine DEKAY.
 \\
  Return type: \textit{void} \\
  Parameters: \textit{none}
\end{description}

\begin{description}
\item[INIPHY]  Initializes parameters of QED, common block QEDPRM, \\
  Return type: \textit{void} \\
  Parameters:
  \begin{enumerate}
    \item \textit {float PI}=3.1415...
    \item \textit {float ALFINV }=1/$\alpha_{QED}$
    \item \textit {float ALFPI }=$\alpha_{QED}/\pi$
    \item \textit {float XK0} = dummy at present
  \end{enumerate}
\end{description}

\begin{description}
\item[DEXAY] Generates decay of a polarized  $\tau$. DEKAY is more powerful for
  spin effects and should be used in preference to this.\\
  Return type: \textit{void} \\
  Parameters:
  \begin{enumerate}
    \item \textit {int state} parameter defining the choice between $\tau^+$ and $\tau^-$. If set to 1, a decay of $\tau$
with ID=IDFF will be performed, otherwise it will be for ID=-IDFF. In both cases FILHEP will be invoked to store 
the appropriate decay products in the event record.

    \item \textit {double pol[4]} input $\tau$ polarization vector.
  \end{enumerate}
\end{description}

\begin{description}
\item[DEKAY] Generates an unpolarized $\tau$ decay  \\
  Return type: \textit{void} \\
  Parameters:
  \begin{enumerate}
    \item \textit {int state} parameter defining choice between $\tau^+$ and $\tau^-$.
If set to 1, decay of a $\tau$
with ID=IDFF will be performed, for 2 it will be with ID=-IDFF. For 11 and 12  FILHEP will be invoked to store 
the appropriate decay products in the event record.
                  
    \item \textit {double [4]} returns vector H see Section \ref{dwaTrzy} 
  \end{enumerate}
\end{description}

\subsubsection{ C++ Routines Called by {\tt TAUOLA FORTRAN}   }

\begin{description}
\item[void filhep\_] {\bf (int n, int status, int pdg\_id, int mother\_first, 
  int mother\_last, int daughter\_first, int daughter\_last, float p4[4], 
  float p\_inv\_mass, bool photos\_flag)} puts the particle into the event record. A long list of 
its parameters (variables named in {\tt HEPEVT} style), is given  below:

  \begin{enumerate}
    \item \textit {n} Index of the particle
    \item \textit {status} Status code of the particle
    \item \textit {pdg\_id} PDG id of the particle
    \item \textit {mother\_first} Index to the particles first mother
    \item \textit {mother\_last} Index to the particles last mother
    \item \textit {daughter\_first} Index to the particles first daughter
    \item \textit {daughter\_last} Index to the particles last daughter
    \item \textit {p4[4]} 4-momentum, the last component is energy 
    \item \textit {p\_inv\_mass} Mass of the particle
    \item \textit {photos\_flag} Should {\tt PHOTOS} be called for this particle
  \end{enumerate}
\end{description}

\begin{description}
\item[void tralo4\_] (float * kto, float p[4], float q[4], float * ams);
FORTRAN routine which is used to boost the four vector p[4] from first/second
$\tau$'s ({\tt kto}=1/2)  rest frame to laboratory frame q[4]. ams denotes a four 
vector mass or virtuality.
\end{description}

\begin{description}
\item[float amas4\_](float*); returns the mass of the argument four vector.
\end{description}

\begin{description}
\item[void bostr3\_](float*, float*, float*); This routine performs boosting 
(with boost parameter given by the first argument) of a four vector (second argument)
into a four vector (third argument). 
\end{description}

\section{Appendix: User Guide}
\label{sec:User Guide}
\subsection{Installation}
\label{sec:Installation}
The main interface library requires that {\tt HepMC} \cite{Dobbs:2001ck} (version 2.04 or later)  has been installed
and its location has been provided during the configuration step. This is sufficient to compile the interface and to run the simple, standalone example.

However, in order to run more advanced examples located in the {\tt /examples} directory, it is required to install also:

\begin{itemize}
  \item {\tt ROOT} \cite{root-install-www} version 5.18 or later
  \item {\tt PYTHIA 8.1} \cite{Sjostrand:2007gs} or later. {\tt PYTHIA 8.1} must be compiled with {\tt HepMC} 2.xx
        so that the {\tt PYTHIA} library {\tt hepmcinterface} exists.
  \item {\tt MC-TESTER} \cite{Golonka:2002rz,Davidson:2008ma} version 1.24 or later.
        Do not forget to type {\tt make libHepMCEvent} after compilation of {\tt MC-TESTER} is done.
\end{itemize}

In order to compile the {\tt TAUOLA C++ Interface}:
\begin{itemize}
 \item Execute {\tt ./configure} with the additional command line options:
   \subitem {\tt --with-HepMC=$<$path$>$} provides the path to the {\tt HepMC} installation directory. One can also set the {\tt HEPMCLOCATION} variable instead of using this directive. This path is required for the interface to compile.
   \subitem {\tt --prefix=$<$path$>$} provides the installation path. The {\tt include} and {\tt lib} directories will be copied there if {\tt make install} is executed later. If none has been provided, the default directory for installation is {\tt /usr/local}.
 \item Execute {\tt make}
 \item Optionally, execute {\tt make install} to copy files to the directory provided during configuration.
\end{itemize}

After compiling the {\tt TAUOLA/tauola-fortran} part, the {\tt TAUOLA C++ interface} will be compiled and the {\tt /lib} and {\tt /include} directories will contain the appropriate libraries and include files.

In order to compile the examples, enter the {\tt /examples} directory and:
\begin{itemize}
  \item Compile {\tt TAUOLA C++ interface}
  \item Execute {\tt ./configure} to determine which examples can be compiled. Additional paths can be provided as command line options:
   \subitem {\tt --with-Pythia8=$<$path$>$} provides the path to the {\tt Pythia8} installation directory. One can set the {\tt PYTHIALOCATION} variable instead of using this directive. This path is required for all additional examples and tests.
   \subitem {\tt --with-MC-Tester=$<$path$>$} provides the path to the {\tt MC-TESTER} installation directory (the {\tt libHepMCEvent} must be compiled as well, see \cite{Davidson:2008ma} for more details). One can set the {\tt MCTESTERLOCATION} variable instead of using this directive. This path is required for all additional examples and tests.  This option implies that {\tt ROOT} has already been installed (since it is required by {\tt MC-TESTER}). The {\tt ROOT} directory {\tt bin} should be listed in   variable {\tt PATH}
and {\tt ROOT} libraries in {\tt LD\_LIBRARY\_PATH}.

  \item execute {\tt make}
\end{itemize}

If neither {\tt Pythia8} nor {\tt MC-TESTER} are present, only the simple example will be provided. The {\tt /examples} directory will contain the compiled example files.

\subsection{Elementary Tests}
\label{sect:elem}
The most basic test which should be performed is verification that the interface is installed correctly, that all
$\tau$ leptons are indeed decayed by the program and that energy 
momentum conservation is preserved. {\tt TAUOLA} has its own database of parameters
and as a consequence the $\tau$ lepton mass may differ between the 
program performing a $\tau$'s production and {\tt TAUOLA} performing its decay. This leads to the sum of 
$\tau$ decay product momenta not exactly matching the $\tau$'s momentum. Although
this effect may seem negligible, it may break numerical stability of 
programs like {\tt PHOTOS} if they are applied later.

Once correct execution of the basic program steps have been confirmed, ie. $\tau$ leptons are 
decayed, energy momentum is conserved and there are no double 
decay occurrences in the event tree, step one of the program installation tests 
 is completed\footnote{
We have  performed such  tests for all choices of the {\tt HepMC} event record obtained 
from  {\tt PYTHIA 8.1} processes and 
listed later in the paper. Further  options for initializations 
(parton shower hadronization or QED bremsstrahlung on/off etc.) were studied.
This installation step  was a necessary one of program development as well.}.

In principle, these tests have to be performed for any new hard 
process and after any new installation. This is to ensure that 
information is passed from the event record to the interface 
correctly and that physics information is filled into {\tt HepMC} 
in expected manner. Misinterpretation of the event record content may result in 
faulty generation by {\tt TAUOLA}. For example spin correlations may be 
missing or badly represented, or some $\tau$ leptons may remain 
undecayed.

\subsection{Executing Examples}
Once elementary tests are completed one can turn to the more advanced ones.
The purpose  is
not only to validate the installation but  to demonstrate how the
interface can be used and how spin affects some distributions.

The examples can be run  by executing the appropriate {\tt .exe} file in the {\tt /examples} directory.
In order to run some more specific tests for spin effects and decays of the following intermediate states:
 $Z$, $W$, $H$, $H^\pm$, the main programs residing in subdirectories of the same name placed in the {\tt /examples/testing} directory
should be executed. For tests of all $\tau$  decay modes directory {\tt /examples/testing/tau} is prepared.
In all cases the following actions have to be performed:

\begin{itemize}
  \item Compile {\tt TAUOLA C++ Interface} as well as the examples. 
 \item  Check that the appropriate system variables are set: normally set by the script \\
{\tt /configure.paths.sh} (the configuration step mentions this script).
  \item  Enter the {\tt /examples/testing} directory. Modify test.inc if needed.
  \item enter the chosen directory and execute {\tt make}.
\end{itemize}

The appropriate .root files as well as .pdf files generated by {\tt MC-TESTER} will be created inside the chosen directory. 
One can execute 'make clobber' to clean the directory. 
One can also execute 'make' inside the {\tt /examples/testing} directory to run all available tests one after another.
New source
code changes can easily be validated in this way.
Tests are run using {\tt examples/taumain\_pythia\_example.exe} 
and booklets will be produced with comparisons to the benchmark files.

A set of benchmark {\tt MC-TESTER} root files are packed with the interface
distribution in the subdirectories of {\tt examples/testing/}. They can be used
as examples to start new work or simply to construct comparison
plots to validate new versions or new installations of {\tt TAUOLA Interface}.

In Appendix \ref{sec:User Configuration} possible  modifications to the  
examples settings 
are discussed. This may be interesting as an initial step for users 
physics studies.  Numerical results of some of these tests
are collected in Section \ref{sec:results}
and can be thus reproduced by the user.

\subsubsection{Monitoring $\tau$ Decay  Channels}

It is important to check, if  the $\tau$ decays themselves,
are generated correctly on the user platform.
For that purpose, our last demo (directory {\tt /examples/testing/tau}) is prepared.  If the test is activated,
the user performs
a standard {\tt MC-TESTER} comparison of his program execution with the  pre-generated one (of 10 million events).
In this case all $\tau$ decay modes are activated and {\tt MC-TESTER} is simply
analyzing $\tau$ decays themselves.

\subsection{Library Linking}
In order to link the libraries to the user's project, both the static libraries and shared objects are
constructed. To use {\tt TAUOLA FORTRAN} and {\tt TAUOLA Interface} in an external project additional 
compilation directives are required. For the static libraries:
\begin{itemize}
  \item add {\tt -I<TauolaLocation>/include } at the compilation step,
  \item add {\tt <TauolaLocation>/lib/libTauolaCxxInterface.a } and 
\\ {\tt <TauolaLocation>/lib/libTauolaFortran.a } at the linking step.
\end{itemize}
For the shared objects:
\begin{itemize}
  \item add {\tt -I<TauolaLocation>/include } at the compilation step,
  \item add {\tt -L<TauolaLocation>/lib } along with {\tt -lTauolaCxxInterface -lTauolaFortran } at the linking step.
  \item  {\tt TAUOLA} libraries must be provided for the executable; eg. with the help of  {\tt LD\_LIBRARY\_PATH }.
\end{itemize}
The {\tt <TauolaLocation>} denotes the  path to {\tt TAUOLA} installation directory. 

\subsection{ Known Issues}
\label{subs:Issues}

We list  here difficulties  we've encountered during the testing phase and during 
installation for particular configurations.

The first problem occurs if a user incorrectly configures the units of
{\tt PYTHIA} to be different from the units in {\tt HepMC}. (for
example: if {\tt PYTHIA} produces output in MeV, while {\tt HepMC}
interprets input as being in GeV).  In this case, the
built-in routines of {\tt TAUOLA Interface} will treat the input as being in
GeV and will adapt its output to those units as well.  If this kind of
situation occurs (the user will be notified by many warnings of
four-vector momentum not being conserved), one can force {\tt HepMC}
to use MeV units just before filling it with the Pythia event record
data. The {\tt HepMC} event will be automatically converted to GeV
when {\tt TAUOLA Interface} is called.

Another example of a known  compatibility issue arose because of a difference between the assumed default 
   {\tt HepMC} version 2.05 and  version 2.03 (currently used, for example, in Athena%
\footnote{Software framework   of the ATLAS collaboration.}). 
In this case, the script  {\tt platform/to-HepMC-2.03.sh} will be automatically invoked during the configuration
step.
However, in version 2.03 methods for unit conversion
are absent,  therefore GeV and mm will be expected for input and used for output. The method
  {\tt Tauola::setUnits(\dots)},
described in Appendix~\ref{subsection:tauola:other_methods}, becomes dummy.

At present, modification to our {\tt C++ Interface}  has to be introduced
eg.  for use
in the {\tt Athena} system
of the ATLAS collaboration software.  This is to allow for backwards compatibility with 
older versions of
{\tt HepMC}  and to prevent name clashes with the old {\tt TAUOLA FORTRAN
Interface} in environments where both interfaces are loaded concurrently.
 On the other 
hand, there is no problem with the library {\tt /lib/libTauolaFortran.a } of the 
main part of the {\tt TAUOLA FORTRAN} code itself. The version used by {\tt Athena} can be loaded 
instead of ours. In 
{\tt Athena}, the {\tt binp} variant of the {\tt cleo} initialization is used for
{\tt TAUOLA}; in this variant, 
 for the $4\pi$  decay modes of $\tau$'s, parametrization based 
on Novosibirsk data 
is used \cite{Bondar:2002mw}. 

All necessary changes for our {\tt C++ TAUOLA Interface} can be introduced 
with use of the script {\tt platform/to-Athena.sh}.
It can be invoked by 
executing the {\tt make athena} command in the main directory. 
Recompilation of the interface must then be performed.

\section{Appendix: User  Configuration}
\label{sec:User Configuration}
In this section we give a description of how the user can configure {\tt TAUOLA FORTRAN }
and the {\tt TAUOLA Interface}.
All configuration is done via the static class {\tt Tauola}. Below is the 
complete list of user configurable parameters and basic information on their
meaning.

\subsection{Spin Correlation}
By default, all spin correlations are turned on. However one may be interested
to  partially or completely switch off their effects for the sake of numerical 
experiments which validate whether a measurement will be sensitive to certain spin correlation 
components.
This technique may be useful to evaluate the significance of spin correlations for 
signal/background separation as well. 

Several partial treatments of spin correlations
are possible. In general, the most complete   intervention is to simply rewrite the matrix $R_{ij}$ 
for the particular channel. The following methods are nonetheless provided:

\begin{itemize}
 \item {\tt Tauola::spin\_correlation.setAll(bool flag)} \hfill \\
Turns all spin correlation computations on or off depending on the flag, which can be either \textbf{true} or \textbf{false}.
Note: this should be called after Tauola::initialise().
 \item {\tt Tauola::spin\_correlation.HIGGS=flag} \hfill \\
Turns particular spin correlation computation on or off for a given $\tau$ parent depending on the flag which can be either
\textbf{true} or \textbf{false}. 
Implementation of this switch is provided for: \textbf{GAMMA, Z0, HIGGS, HIGGS\_H, HIGGS\_A, 
HIGGS\_PLUS, HIGGS\_MINUS, W\_PLUS, W\_MINUS}. The keywords denotes the $\tau$ parent.
\end{itemize}

\noindent
 \textbf{Example:} 

\noindent
{\tt Tauola::spin\_correlation.setAll(false); } \\
{\tt Tauola::spin\_correlation.HIGGS=true; } \\
{Turns all spin correlations off, except HIGGS.} \hfill \\

 Finally one can replace density matrix following description given in Appendix~\ref{sect:Importing}
also in this case one  does not need to recompile of the code.

\subsection{Decay Mode Selection}
By default, all $\tau$ decay modes will be generated according to predefined branching fractions.  
Methods to modify the default values are provided:

\begin{itemize}
 \item {\tt Tauola::setSameParticleDecayMode(int mode)} \hfill \\
Set the decay mode of the  $\tau$ with the same PGD code as set in \\
Tauola::setDecayingParticle() (by default this sets the decay mode of $\tau^-$). \\
 \item {\tt Tauola::setOppositeParticleDecayMode(int mode)} \hfill \\ 
Set decay mode of the  $\tau$ with the opposite PGD code as set in \\
Tauola::setDecayingParticle() (by default this sets the decay mode of $\tau^+$). \\ \\
\noindent
\textbf{Example:} \hfill \\
{\tt Tauola::setSameParticleDecayMode(Tauola::PionMode); } \\
{\tt Tauola::setOppositeParticleDecayMode(4); } \\
\emph{ Forces only the modes $\tau^- \rightarrow \pi^- \nu_{\tau}$ and 
  $\tau^+ \rightarrow \rho^+ \nu_{\tau} (\rho^+ \rightarrow \pi^+ \pi^0)$ to be generated  }\hfill \\ 
 \item {\tt Tauola::setTauBr(int mode, double br)} \hfill \\
Change  the $\tau$ branching ratio for channel
{\it mode} from default to {\it br}. Note: this should be called after Tauola::initialise(). \hfill \\ 
\noindent
\textbf{Example:} \hfill \\
{\tt Tauola::setTauBr(3, 2.5)};\\
\emph{Sets rate for channel $\tau^\pm \rightarrow \pi^\pm \nu_{\tau}$ to 2.5. All channel rates
may not sum to unity, normalization
will be perfrmed anyway. }\hfill \\
\item The  {\tt int mode} enumerators which are arguments 
  of {\tt setOppositeParticleDecayMode}, {\tt setSameParticleDecayMode}, 
  {\tt setTauBr} have the following meaning:
      \begin{itemize}
        \item 0 - {\tt Tauola::All} - All modes switched on
        \item 1 - {\tt Tauola::ElectronMode} - $\tau^{\pm} \rightarrow e^{\pm} \nu_{\tau} \nu_{e}$
        \item 2 - {\tt Tauola::MuonMode} - $\tau^{\pm} \rightarrow \mu^{\pm} \nu_{\tau} \nu_{\mu}$
        \item 3 - {\tt Tauola::PionMode} - $\tau^{\pm} \rightarrow \pi^{\pm} \nu$
        \item 4 - {\tt Tauola::RhoMode} - $\tau^{\pm} \rightarrow \rho^{\pm} \nu$
        \item 5 - {\tt Tauola::A1Mode} - $\tau^{\pm} \rightarrow A_1^{\pm} \nu$
        \item 6 - {\tt Tauola::KMode} - $\tau^{\pm} \rightarrow K^{\pm} \nu$
        \item 7 - {\tt Tauola::KStarMode} - $\tau^{\pm} \rightarrow K^{*\pm} \nu$
        \item 8 - $\tau^{\pm} \rightarrow 2 \pi^{\pm}\pi^{\mp}\pi^0 \nu$
        \item 9 - $\tau^{\pm} \rightarrow 3 \pi^0 \pi^{\pm} \nu$
        \item 10 - $\tau^{\pm} \rightarrow 2 \pi^{\pm}\pi^{\mp}2\pi^0 \nu$
        \item 11 - $\tau^{\pm} \rightarrow 3 \pi^{\pm} 2\pi^{\mp}  \nu$
        \item 12 - $\tau^{\pm} \rightarrow 3 \pi^{\pm} 2\pi^{\mp}\pi^0 \nu$
        \item 13 - $\tau^{\pm} \rightarrow 2 \pi^{\pm}\pi^{\mp}3\pi^0 \nu$
        \item 14 - $\tau^{\pm} \rightarrow K^{\pm} K^{\mp} \pi^{\pm} \nu$
        \item 15 - $\tau^{\pm} \rightarrow K^0 \bar{K^0}\pi^{\pm} \nu$
        \item 16 - $\tau^{\pm} \rightarrow K^{\pm}K^0\pi^0 \nu$
        \item 17 - $\tau^{\pm} \rightarrow 2\pi^0K^{\pm} \nu$
        \item 18 - $\tau^{\pm} \rightarrow \pi^{\pm}\pi^{\mp} K^{\pm} \nu$
        \item 19 - $\tau^{\pm} \rightarrow \pi^{\pm}\pi^0 \bar{K^0} \nu$
        \item 20 - $\tau^{\pm} \rightarrow \eta \pi^{\pm} \pi^0 \nu$
        \item 21 - $\tau^{\pm} \rightarrow \pi^{\pm}\pi^0 \gamma \nu$
        \item 22 - $\tau^{\pm} \rightarrow K^{\pm}K^0 \nu$
      \end{itemize}
 \item {\tt Tauola::setTaukle(double  bra1, double brk0, double brk0b,double brks)} \hfill \\
Change  the $\tau$ sub channels branching ratio between (i) $a_0 \to \pi^+\pi^+\pi^-$ and $a_0 \to \pi^0\pi^0\pi^+$
(ii) subchannels of $K_0$ (iii)  subchannels of $\bar K_0$ and (iv) subchannels of $ K^*$.
Note: this should be called after Tauola::initialise().

\noindent
\textbf{Example:} \hfill \\
 {\tt Tauola::setTaukle(0.5, 0.5, 0.5, 0.6667);} \\
\emph{ Set the parameters to their default values}\hfill \\

 \hfill \\
\end{itemize}

\subsection{Decaying Particle}

The following method is prepared to impose the sign for the 'first $\tau$', that is
to reverse signs of  {\tt SameParticle} and {\tt OppositeParticle} $\tau$:
\begin{itemize}
\item {\tt Tauola::setDecayingParticle(int pdg\_id) } \hfill \\
  Set the PDG id of the particle which {\tt TAUOLA} should decay as 'first $\tau$'.
  Both particles with pdg\_id and -1*pdg\_id will be decayed. 
  Default is 15, one may want to use -15 for special applications.\\
\end{itemize}

\noindent \textbf{Example:} \hfill \\
  {\tt Tauola::setDecayingParticle(-15); } \\
\emph{Set {\tt SameParticle} $\tau$ to be $\tau^+$}
\hfill \\

\subsection{Radiative Corrections}
The user may want to configure parameters used in the generation of QED corrections in the 
leptonic decay channels of 
$\tau$s. For that purpose the following methods are provided:
\begin{itemize}
\item {\tt Tauola::setRadiation(bool switch) } \hfill \\
Radiative corrections for leptonic $\tau$ decays may be switched on or off
by setting the switch to \textbf{true} or \textbf{false} respectively.
By default this is \textbf{true}\footnote{Only in the case of leptonic $\tau$ decays
can radiative corrections be generated in {\tt TAUOLA} \cite{Jezabek:1991qp}.
The algorithm relies on the first order complete matrix element and no 
exponentiation is available. If the multiple photon option is requested or if 
radiative corrections for other decay channels are needed
{\tt PHOTOS} Monte Carlo can be used instead \cite{photosC++}. In \cite{Was:2004dg} it was shown, that the numerical effects due to the parts not included in {\tt PHOTOS} of the first order matrix element is numerically more significant than multiple photon effects. This conclusion is based on our standard numerical tests and will not necessarily be the case for other applications.}.
\item {\tt Tauola::setRadiationCutOff(double cut\_off) } \hfill \\
Set the cut-off for radiative corrections of $\tau$ decays. 
The default of 0.01 means that only photon of energy (in its rest frame) up to 0.01 of half of the decaying
particle mass will be explicitly generated.
\end{itemize}   

\noindent \textbf{Example:} \hfill \\
  {\tt Tauola::setRadiation(false); } \\
  \emph{Switch radiative corrections off in $\tau$ decays } \hfill \\

\subsection{Decay of Final State Scalars}
In some cases a user may want {\tt TAUOLA} to decay short living scalar particles
produced in $\tau^\pm$ decays, rather than invoking a host generator for the post processing 
step.
For that purpose a special algorithm is prepared, even though high precision
is then not assured. This might not be a problem if the algorithm is used for $\tau$ decays 
only where 
events with such decays are rather rare:

\begin{itemize}
 \item {\tt Tauola::setEtaK0sPi(int a, int b, int c) } \hfill \\ The
   three parameters $a$, $b$ and $c$ switch on or off the decay of
   $\eta$, $K_s^0$ and $\pi^0$ respectively. A value of 1 is on and 0 is off.
\end{itemize}
 
\noindent
\textbf{Example:} \hfill \\
  {\tt Tauola::setEtaK0sPi(1,0,1); } \\
\emph{In event branch starting from $\tau$, $\eta$ and $\pi^0$ decay, 
but $K_s^0$ remains undecayed.
 } \hfill \\

\subsection{Scalar-Pseudoscalar Higgs}
Users may wish to study spin correlations in processes involving scalar, pseudoscalar
or mixed scalar-pseduoscalar decays into $\tau$'s. All
options are supported by this interface.
The spin density matrix will be calculated correctly for scalar Higgs
(assumed PDG id of 25) and for pseudoscalar Higgs (assumed PDG id of 36)
without any additional user configuration. For other cases, such as a mixed 
scalar-pseduoscalar Higgs or the decay of non-Higgs scalar particles, 
the following methods are provided:

\begin{itemize}
\item {\tt Tauola::setHiggsScalarPseudoscalarMixingPDG(int pdg\_code)} \hfill \\
The PDG Monte-Carlo code of the Higgs which should be treated by the interface as
a scalar-pseudosclar mix. The default value is PDG id 35. Please note that if
{\emph pdg\_code} is set to the value of an existing spin case (eg. 25, the 
regular scalar Higgs) the scalar-pseudoscalar case will be assumed.
\item {\tt Tauola::setHiggsScalarPseudoscalarMixingAngle(double angle)} \hfill \\
The scalar-pseudoscalar mixing angle. ie. $\phi$ in the coupling: 
$\bar{\tau}(cos(\phi)+isin(\phi)\gamma_5)\tau$. By default $\phi=\frac{\pi}{4}$.\end{itemize}
 
\noindent \textbf{Example:} \hfill \\
  {\tt Tauola::setHiggsScalarPseudoscalarMixingPDG(24); } \\
  {\tt Tauola::setHiggsScalarPseudoscalarMixingAngle(3.1415/3.0); } \\
  {Spin correlations will be calculated for the Higgs boson as though it is a scalar-pseduoscalar with mixing angle of $\frac{\pi}{3}$ \hfill \\

\subsection{Helicity States and Electroweak Correcting Weight}
\label{subsection:eHelicity-ew}

Independent of the generation process,
the information on helicities of $\tau^+$ and $\tau^-$ 
can be returned\footnote{ One has to be careful because the actual sign may depend on the process
and boosting routine.}  with the help of accessors:

\begin{itemize}
  \item {\tt int Tauola::getHelPlus()}
  \item {\tt int Tauola::getHelMinus()}
\end{itemize}

Note that these helicities are not used in the interface and carry approximate information only.

The electroweak weight can be returned with the help of accessors:
\begin{itemize}
  \item {\tt double Tauola::getEWwt()} - for cross section with electroweak corrections included
  \item {\tt double Tauola::getEWwt0()} - for cross section at born level
\end{itemize}

These methods provide information once processing of a given event is completed.

\subsection{Use of {\tt TAUOLA decayOne} method}
\label{TAUOLAgun}

In Section~\ref{sect:Outline} an algorithm to decay all $\tau$ leptons present in the event 
record is explained. For that purpose  {\tt decayTaus()} method is provided.
To decay a single $\tau$ lepton in a way independent of the
event record content another, simple method is provided. 
Obvious examples when it can be useful, 
are processes where the hard matrix element originates from models
of new physics, and different flavours of such models are to be tested.
  In such cases,
universal methods of finding spin states of the $\tau$ to be decayed may not exist. 
 Depending of the precision required one may need to: decay 
a $\tau$ without taking into account its spin state, impose its individual 
spin state as input information or provide a method which can be used for full density matrix 
generation. In the last case control over Lorentz transformations between 
the $\tau$ rest-frame and laboratory frame have to be available for the  
user.

Fortunately for all these applications a rather simple method is sufficient.
 It can be used to generate a decay of
an individual $\tau$, without information on its parents.
\begin{itemize}
  \item {\tt Tauola::decayOne( \\ TauolaParticle *tau, bool undecay, double polx, double poly, double polz) } \hfill \\
	The main routine for decaying the tau. Only the first parameter is mandatory. The first parameter is a pointer to the $\tau$
         that needs to be decayed. \\
	The {\tt undecay} flag determines the reaction that should be taken if the $\tau$ already has daughters. 
        By default the flag is set to {\bf false}, which means that already decayed $\tau$ will be left unchanged. 
        Setting this flag to {\bf true} allows the interface to first undecay the $\tau$ and replace it with a new decay.\\
	The last three parameters are the components of the $\tau$ polarization 3-vector. In the case of {\tt TAUOLA decayOne}, 
        the decayed $\tau$ is treated as a standalone particle, without considering its mothers, daughters or 
        siblings. In case the user wants to input the polarization vector (at default $\tau$ is treated as
not polarized), 
the last three parameters have to be used.
  \item {\tt Tauola::setBoostRoutine( \\ void (*boost)(TauolaParticle *tau, TauolaParticle *target) ) } \\
        Once executed, {\tt Tauola:decayOne}  will use the user function instead of the default one, to boost $\tau$ decay 
        products from their rest frame to the lab frame. Such feature may be essential, in future, 
        for use of {\tt Tauola::decayOne} as part of a user algorithm for 
generation of exact spin correlations  in multi $\tau$ final states.

\end{itemize}

The {\tt single\_tau\_gun\_example.c} is provided in the directory {\tt /examples}. If polarization 
{\tt polx=0, poly=0, polz=1} is chosen, then the helicity state is taken:
left handed $\tau^-$ or right handed $\tau^+$.  If, again as given in the example  {\tt Tauola::setBoostRoutine} is used
with the proposed method, then {\tt polx=0, poly=0, polz=1} will not mean helicity state,
but rather the $\tau$ spin polarization vector oriented along the z axis of the lab frame (in fact along
its space component in the $\tau$ rest-frame). 
Obviously
spin effect chosen this way, will depend on the direction of the $\tau$ momentum.

\subsection{Logging and Debugging}
This section describes the basic functionality of the logging and debugging tool.
For details on its content
 we adress the reader  to comments in the {\tt /src/utilities/Log.h} header file.

Let us present however some general scheme of the tool functioning.
{\tt TAUOLA Interface} allows filtering out some amount of data displayed during the program run
and provides a basic tool for memory leak tracking. The following functions
can be used from within the user program after including the {\tt Log.h} file:
\begin{itemize}
  \item {\tt Log::Summary() } - Displays a summary of all messages.
  \item {\tt Log::SummaryAtExit() } - Displays the summary at the end of a program run.
  \item {\tt Log::LogInfo(bool flag) } \\
        {\tt Log::LogWarning(bool flag) } \\
        {\tt Log::LogError(bool flag) } \\
        {\tt Log::LogDebug(int s, int e) } \\
        {\tt Log::LogAll(bool flag)} \\
        Turns logging of info, warning, error and debug messages on and off depending
        on the flag being true or false. In the case of debug messages - the range of codes
        to be displayed must be provided. By default, only debug messages
        (from 0 to 65535) are turned off. If the range is negative ($s>e$) the debug messages
        won't be displayed. The last option turns displaying all of the above messages on and off.
\end{itemize}

The memory leak tracking function allows checking of whether all memory allocated within {\tt TAUOLA Interface}
 is properly released. However, using the debug option significantly increases the amount of time needed for 
each run. Its  use is therefore recommended  for debugging purposes only. In order to use this option
 modify {\tt make.inc} in the main directory by adding the line: \\
\\ {\tt DEBUG = -D"\_LOG\_DEBUG\_MODE\_" } \\ \\
Recompile the interface.
Now, whenever the program is executed 
a table will be printed at the end of the run, listing all pointers that were not freed, along with the memory they consumed.
If the interface
works correctly without any memory leaks, one should get an empty table.

It is possible to use this tool within the user program, however there are a few limitations.
The debugging macro from "Log.h" can create compilation errors if one compiles
it along with software which has its own memory management system (e.g. {\tt ROOT}).
To make the macro work within a user's program, ensure that {\tt Log.h} is the last header file
included in the main program.
It is enough to  compile the program with the {\tt -D"\_LOG\_DEBUG\_MODE\_" } directive added,
or {\tt \#define \_LOG\_DEBUG\_MODE\_ } placed within the program before include
of the {\tt Log.h} file%
\footnote{Note that {\tt Log.h } does not need to be included within
the user program  for the memory leak tracking tool to be used only for {\tt TAUOLA Interface}.
}.

\subsection{Plots for Debugging and Monitoring}
This section describes the basic functionality of the plotting tool.
Detailed explanations are given in the {\tt /src/utilities/Plot.h} and {\tt /src/utilities/Plot.cxx} files.

The {\tt Plot} class allows generation of data for several plots we use to monitor the 
interface. At present, $\tau$ polarization, as taken from the {\tt SANC} library and used by the interface (including  its
interpolation algorithm) is monitored in this way.
The program generates data files during execution to be used later for graphic output.
This code is not expected to be of a large interest to users. It is mainly for  testing and 
debugging purposes, but may be of interest for installation tests as well.

In order to generate input data for plotting, a few methods have been provided which can be accessed after adding 
{\tt \#include "Plots.h"} in the main program file:
\begin{itemize}
  \item {\tt Plots::setSancVariables(int flavor,double cosTheta) } - sets the variables used by 
the first two tests. The first one is the flavor of incoming particle (by default it's set 
to 1), and the second is value of $cos(\theta)$ (by default - 0).
  \item {\tt Plots::addSancPlot(int i) } - adds the i$^{th}$ test plot of the four tests prepared for  {\tt SANC} 
tables. All or only some of these tests can be run simultaneously. Figures \ref{fig:sigmaUp}, \ref{fig:sigmaDown},
and \ref{fig:polUp}, \ref{fig:polDown} can be in particular reproduced with this method.
  \item {\tt Plots::exitAfterSancPlots(bool exit); } - if set to {\bf true}, force the program to 
exit after the plots have been created. Otherwise, the program will continue normally 
to its usual end.
\end{itemize}

The files generated with this tool can then be used  to make plots with external scripts. 
For this purpose an {\tt /examples/draw.C} ROOT script has been provided. If  {\tt root draw.C}
is executed, 
it checks, by name,  which input data files are present. For existing data files plots are drawn. 
Since generated files contain the test data only,  without much of  explanation of their meaning, 
to interpret them one need to look for a description inside the Plot class source files. 

\subsection{Other User Configuration Methods}
\label{subsection:tauola:other_methods}

The following auxiliary methods are prepared. They are useful for initialization 
or are introduced for backward compatibility.
\begin{itemize}
  \item {\tt Tauola::setUnits(Tauola::MomentumUnits,Tauola::LengthUnits)} \\
        Set output units for momentum (Tauola::GEV / Tauola::MEV)
        and decay vertex position (Tauola::MM / Tauola::CM). Methods are only available for {\tt HepMC 2.04} 
        or higher.
  \item {\tt Tauola::setTauLifetime(0.08711)} \\
        Set the mean $\tau$ lepton lifetime in mm. If the user wants a vertex position
        to be generated by his own method, then the numerical value
        of the $\tau$ lifetime should be set to 0.0 here.
  \item {\tt Tauola::setInitialisePhy(double iniphy\_param)} \\
        Initializes some constants related to QED corrections.
        The variable {\tt iniphy\_param} is at present a dummy. It is prepared to be
        transmitted to some
        old style production code and is kept for backward compatibility.
  \item {\tt Tauola::setRandomGenerator(double (*gen)())} \\
        In {\tt tauola-fortran} the random number generator {\tt RANMAR}
        is used. It is also used in our auxiliary methods which temporarily remain
        in {\tt FORTRAN}. {\tt RANMAR} may need to be replaced or a particular 
        seed has to be chosen. It can be easily done and is explained in
        \cite{Golonka:2003xt}. In the C++ part of the code a user can simply 
        set the pointer to the replacement for an internal random number 
        generator {\tt Tauola::RandomDouble}.
        The generator must return  a {\tt double} from 0 to 1. 
        {\tt Tauola::setRandomGenerator(NULL)} will reset the program back to  
        the default  generator.
\end{itemize}

\section{Appendix: Modifying Electroweak Corrections Module}
\label{section:SANC}

\subsection{SANC Unit Initialization and Input Parameters  }

In this section we describe details of the {\tt SANC} library initialization. Input parameters and constants are collected in several
files located in the directory {\tt /SANC}. The file {\tt  s2n\_declare.h} contains a declaration of all {\tt FORTRAN} variables used 
by the {\tt SANC} NLO Library. The particle masses and coupling constants
are initialized by the {\tt sanc\_input.h} header file. It is called by the {\tt s2n\_init} subroutine 
(see file {\tt s2n\_init.f}), which initializes other parameters like particle mass ratios,
particle charges and weak isospin projection, as well as the value of fictitious photon mass ({\tt thmu2}) used by IR singularity 
regularization and
soft/hard radiation separator ({\tt omega}). Several  user controlled flags are defined:
\begin{itemize}
   \item {\rm \bf iqed} = 1/0 -- NLO QED correction is switched on/off. Default {\rm \bf iqed} = 0
   \item {\rm \bf iew}  = 1/0 -- NLO EW correction is switched on/off. Default  {\rm \bf iew} = 1
   \item {\rm \bf iborn} = 0/1 -- NLO correction are switched on/off; Default {\rm \bf iborn} = 0
   \item {\rm \bf gfscheme} = 1/0 --  calculation schemes: 1 - Fermi Scheme, 0 - Alpha 
Scheme (default).
\end{itemize}
These flags are used to configure table preparation by the  program
{\tt SANC/SANCinterface.cxx}.
For the {\tt SANC} module  structure and its project details, see refs.~\cite{Andonov:2008ga,Andonov:2004hi}.

\subsection{Structure of Files with Pretabulated $R_{ij}$}

In order to generate all pretabulated files the {\tt SANC/SANCinterface.cxx } program is used.
When compiled along with the {\tt SANC} library, {\tt SANC} FORTRAN interface and modules located inside the {\tt SANC/modules } directory,
the program generates two files - {\tt table1-1.txt} and {\tt table2-2.txt} for the quarks down and up respectively.
The program is  invoked with the command {\tt make tables} from the directory {\tt SANC}. 
The third file {\tt table11-11.txt } representing tabulated results  for  electron beams
 will   not be generated automatically. 

The structure of each generated file consists of  several blocks:

\begin{itemize}
  \item Initialization block
    \subitem {\bf Dimensions} - NS1, NS2, NS3 and NCOS values used
             as dimensions of the generated  tables
    \subitem {\bf Ranges} - minimum and maximum values for all three 
             pretabulation ranges of s
  \item Information block:
    \subitem Date and time of generation
    \subitem Full path of the generating program
    \subitem {\tt SANC} library information block
    \subitem {\tt SANC} initialization parameters list
  \item Data block:
    \subitem {\bf BeginRange1} statement
    \subitem tables of NS1 * NCOS  lines for first range
    \subitem {\bf BeginRange2} statement
    \subitem tables of NS2 * NCOS  lines for second range
    \subitem {\bf BeginRange3} statement
    \subitem tables of NS3 * NCOS  lines  for third range
    \subitem {\bf End} statement
\end{itemize}

Lines  within {\bf Data block} consist of 4*4 numbers for $R_{ij}$,
 two extra numbers for the electroweak weight and  an endline character.
Statements used within {\bf Data block} ({\bf BeginRange1, BeginRange2, BeginRange3, End })
are mandatory and must be present in exact form. They mark the beginning and end
of the appropriate data set. The program also checks whether the data block has been
read completely and verifies if variables read at initialization were consistent.

\subsection{Importing SANC Tables into {\tt TAUOLA Interface}}
\label{sect:Importing}
The three files generated by the {\tt SANC} module are loaded into {\tt TAUOLA Interface}
during the initialization step, if they are located in the directory of the main  program. 
When {\tt Tauola::initialize() } is called, the interface searches for the appropriate
files and if they are found - attempts to import the data.

For a file to be loaded correctly, the dimensions of the input file must match the interface settings.
The content and size of the information block is arbitrary. The information of this block will be 
printed only, but not used otherwise.
After reading all the tables from one file, {\tt TAUOLA Interface} checks if the 
end of the data block has been reached and eventually proceeds to the next file.

If the dimensions do not match, the file is inconsistent with the structure 
(the end of the data block has not been reached or the file has insufficient data) 
- it will not be used by {\tt TAUOLA Interface} even if the file was found. 
In that case the default density matrix will be used.
{\tt TAUOLA Interface} will attempt to read all of the three files, but if either one is 
incorrect or missing,  only the data from those files that have been loaded
correctly will be used.

If the need arises to modify the default tables distributed with {\tt TAUOLA}, the {\tt SANC} folder includes all routines
needed to generate new tables along with a {\tt Makefile} with a few options, including:
\begin{itemize}
  \item {\tt make } - used to recompile the library, modules and LoopTools needed for generation.
  \item {\tt make clobber } - might be needed to remove previous compilation.
  \item {\tt make tables } - used to compile the interface code and generate the tables.
\end{itemize}

The C++ interface to the {\tt SANC} library is located in the {\tt SANC/} folder.
{\tt SANCinterface.cxx } represents the main program with setting
of options for table making.
Options for table layout are  explained in comments within the file.
The interface will be recompiled every time {\tt make tables} is used.
 If needed, further
program options, such as initialization of electroweak coupling constants or particle 
masses for
calculation of electroweak corrections can be modified. 
The {\tt SANC/SANCtable.cxx } file consists of routines for actual calculation 
of table entries from  spin amplitudes  calculated from {\tt SANC}.
Changes which may be of interest to advanced users should be done in this file.

\end{document}